\documentclass[11pt,onecolumn,draftcls]{IEEEtran}
\usepackage{amsmath,epsfig,bm}
\usepackage{subfigure}
\newcommand{\y}{\mathbf{y}}
\newcommand{\x}{\mathbf{x}}

\newcommand{\h}{\mathbf{h}}
\newcommand{\w}{\mathbf{w}}
\newcommand{\g}{\mathbf{g}}
\newcommand{\I}{\mathbf{I}}

\newcommand{\X}{\mathbf{X}}
\newcommand{\K}{\mathbf{K}}

\renewcommand{\S}{\mathbf{S}}

\newcommand{\W}{\mathbf{W}}

\newcommand{\G}{\mathbf{G}}
\renewcommand{\H}{\mathbf{H}}

\newcommand{\exE}{\mathbb{E}}

\newcommand{\D}{\mathbf{D}}

\usepackage{pifont}
\usepackage{graphicx}
\usepackage{amssymb}
\usepackage{amsmath}
\usepackage{cancel}
\usepackage[normalem]{ulem}
\usepackage[dvips]{color}
\usepackage{algorithm}  %for code
\newfont{\fsc}{eusm10}                         % frenchscript letters

\newtheorem{corollary}{\noindent Corollary}
\usepackage{lineno}
\modulolinenumbers[1]

\usepackage{algorithmic}
\usepackage{algorithm}

\newcommand{\LINE}{	\item[]}
\title{Channel Tracking for Relay Networks via Adaptive Particle MCMC}

\author{
\IEEEauthorblockN{Ido Nevat$^1$,
                  Gareth W. Peters $^{2,3}$,
                  Arnaud Doucet $^4$
                  and Jinhong Yuan$^5$\\
\IEEEauthorblockA{$^1$
Wireless \& Networking Tech. Lab, CSIRO, Sydney, Australia.}\\
\IEEEauthorblockA{$^2$
School of Mathematics and Statistics, University of NSW, Sydney, Australia.}\\
\IEEEauthorblockA{$^3$
CSIRO Mathematical and Information Sciences, Sydney, Australia.}\\
\IEEEauthorblockA{$^4$
Department of Computer Science, University of British Columbia, Vancouver, Canada}\\
\IEEEauthorblockA{$^5$
School of Electrical Engineering, University of NSW, Sydney, Australia.}} }

\begin{document}
\maketitle
\begin{abstract}
This paper presents a new approach for channel tracking and parameter estimation in cooperative wireless relay networks. We consider a system with multiple relay nodes operating under an amplify and forward relay function.
We develop a novel algorithm to efficiently solve the challenging problem of joint channel tracking and parameters estimation of the Jakes' system model within a mobile wireless relay network. This is based on \textit{particle Markov chain Monte Carlo} (PMCMC) method. In particular, it first involves developing a Bayesian state space model, then estimating the associated high dimensional posterior using an adaptive Markov chain Monte Carlo (MCMC) sampler relying on a proposal built using a Rao-Blackwellised Sequential Monte Carlo (SMC) filter.
\end{abstract}
%%%%%%%%%%%%%%%%%%%%%%%
\section{Introduction}
The relay channel, first introduced by van der Meulen \cite{vandermeulen:1971}, has recently received considerable attention due to its potential in
wireless applications. Relaying techniques have the potential to provide spatial diversity, improve energy efficiency, and
reduce the interference level of wireless channels, see \cite{nosratinia2004ccw}, \cite{laneman:2004} and \cite{laneman2003dst}.

There are a number of issues to be considered when designing a relay network, the more important of these include: the topology
of the relay network; the number of hops in the relay; the number of relays to include in the network; and the type of relaying
function to incorporate, in order to optimise transmission quality of service requirements.

In order to utilise the relay channel, an accurate channel state information (CSI) is required at the destination.
Periodic insertion of known symbols (pilots) along with the transmitted data is widely used for channel estimation. For example, under an Orthogonal Frequency Division Multiplex (OFDM) system, some of the subcarriers are dedicated to pilot symbols \cite{tong2004paw}, \cite{coleri2002cet}, \cite{li2000psa}, \cite{tufvesson1997pac}, \cite{morelli2001cpa}.

All current works in the literature concentrate on performing channel estimation in a static environment, where the channels are assumed to be constant: in \cite{gao2008channel}, the authors designed two linear estimators, namely the Least Squares (LS) and Linear Minimum Mean Square Error (LMMSE)) for Amplify and Forward (AF) based relay networks; in \cite{yan2008low}, an algorithm for LMMSE channel estimation in OFDM based relay systems was derived; and in \cite{behbahani2008channel}, a training based LMMSE channel estimator for time division multiplex AF relay networks was proposed.
%%%%%%%%%%%%%%%%%%%%%%%%%%%%%%%%%%%%%%%%%%%%

%%%%%%%%%%%%%%%%%%%%%%%%%%%%%%%%%%%%%%%%%%%%
The problem of joint channel tracking and parameter estimation for relay wireless links has not been addressed. This involves jointly estimating Jakes' model parameters for each link in the relay system, which we refer to as static parameters throughout the paper, and the non-linear channel tracking problem. This will be performed in a robust statistical estimation framework.
The focus will be on channel tracking in a dual hop relay network in which the number of parallel relays is arbitrary
and the type of relaying function can be general.

The overall channel from the Base Station (BS) to the Mobile Station (MS) via the relay is a cascade of two links: the BS-relay link and the relay-MS link.
Modeling the individual channels can proceed according to Jake's model. This provides a good approximation of the channels dynamics by using a first order Gauss-Markov processes \cite{Chang:1996}, \cite{Komninakis:2002}.

\textbf{Contribution:} in this work, we propose a novel statistical relay model to address the problem of inference and estimation for channel tracking and parameter estimation. We structure the problem of channel tracking such that the overall channel from source to destination is only estimated at the destination. The advantage of this approach is that very general relay networks with varying processing capabilities can be considered in this system. For example, the popular relay functionality of AF can be considered here without any additional computational overhead at the relay nodes, and all the statistical estimation and resulting computational effort is centralised at the BS.
 The statistical signal processing methodology we develop to address this problem utilises a novel development of a recent particle Markov chain Monte Carlo (PMCMC) algorithm \cite{andrieu-particle}. PMCMC allows one to efficiently sample from very high dimensional, strongly correlated multivariate time series models. In the context of channel tracking, these distributions are obtained via filtering recursions, and the challenge is to jointly sample the latent process and static parameters in order to perform estimation.

The motivation of this paper is therefore to provide a system model and estimation procedure for relay network channel estimation. The consequences of this are that with improved channel estimates, tasks such as detection, synchronisation, power allocation and percoding can be improved.

The paper is structured as follows: in Section \ref{ModelSection} a stochastic system model is developed and a Bayesian inference methodology is provided. In Section \ref{ChannelTrackingSamplingFramework} the estimation problem is presented, this involves development of the novel PMCMC sampling methodology and comparison to a standard less efficient approach is discussed.
In Section \ref{complexity} a complexity analysis of the proposed algorithms is provided. Section \ref{simulationresults} provides extensive simulation results firstly investigating algorithmic performance, and then providing detailed study of channel estimation at different SNR values.
Conclusions are provided in Section \ref{conclusions}.

\textbf{Notation:} the notation used throughout this paper will involve: capitals to denote random variables and lower case to denote realizations of random variables; bold face to denote vectors and non-bold for scalars; super script will be used to refer to the index for a particular relay in the network; sub-script will denote discrete time, where $h_{1:T}$ denotes $h_1,\ldots,h_T$; and in the sampling methodology combining MCMC and particle filtering we use the following notation $[\cdot](j,i)$ to denote the $j$-th state of the Markov chain for the $i$-th particle in the particle filter. 
In addition, we denote the proposed Markov chain state by $[\cdot]^*(j,i)$.

\section{SYSTEM DESCRIPTION} \label{ModelSection}
\begin{figure}[h]
    \centering
        \epsfysize=7cm
        \epsfxsize=10cm
        \epsffile{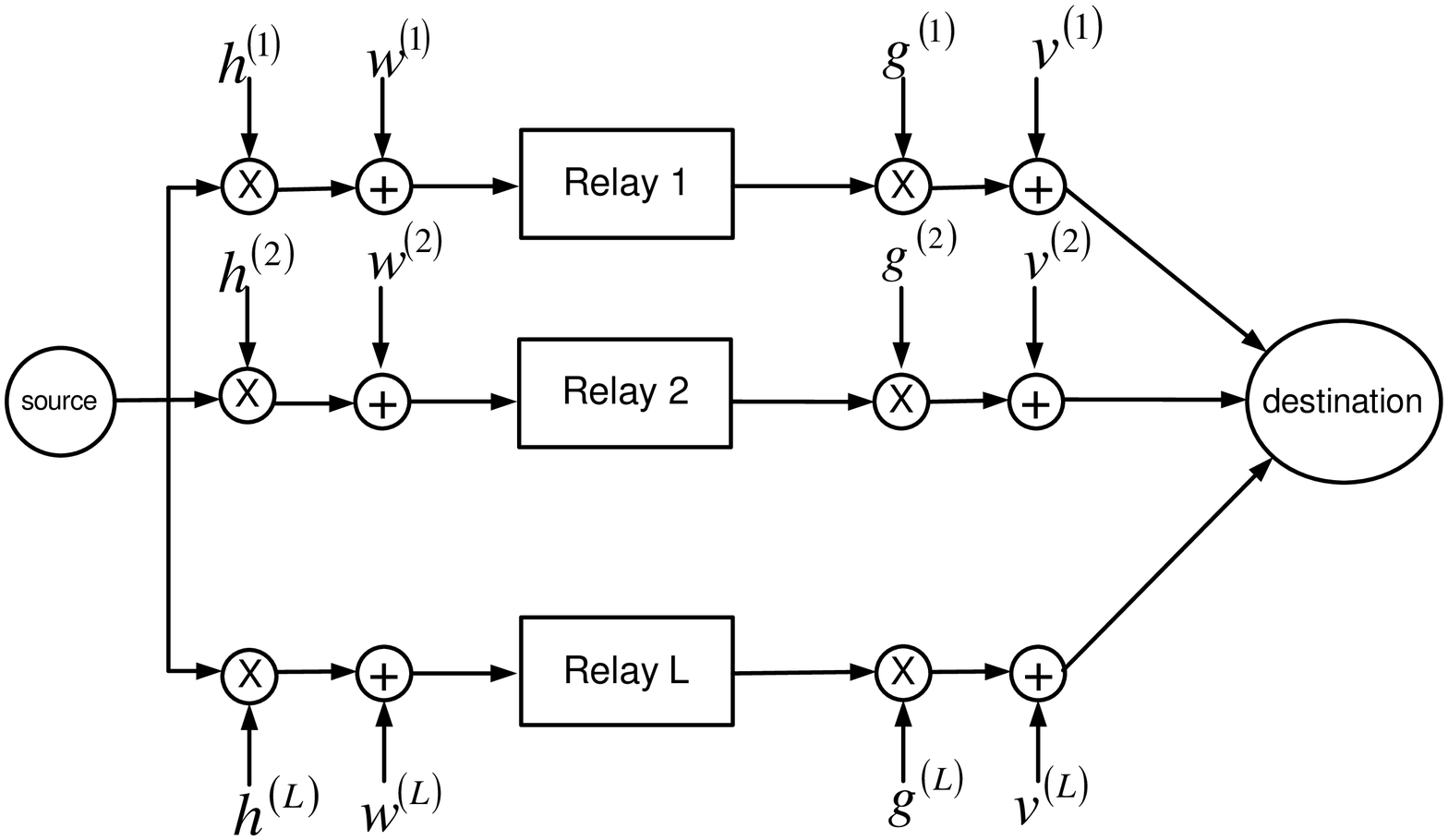}
        \caption{System model of the relay network with $L$ relays and a dual hop}
    \label{fig:system}
\end{figure}

%%%%%%%%%%%%%%%%%%%%%%%%%%%%%%%%%%%%%%%%%%%%%%%%%%%%%%%%%%%%%%%%%%%%%%%%%%%%%%%
The system model considered in this paper is presented in Fig. \ref{fig:system}.
We consider the case where one mobile station is transmitting to a BS via $L$ mobile relay links.
For simplicity, we assume that the number of relay links $L$ is constant, and that during the entire transmission, the mobile station and the relays communicate with the same BS. We consider frequency-flat fading characteristics, for example via the use of OFDM modulation. In a typical OFDM system some of the subcarriers are dedicated for use as pilot symbols, see \cite{tong2004paw}, \cite{coleri2002cet}. We adopt this concept as it allows us to perform the channel tracking on a symbol by symbol basis within a frame. The filtering framework we develop updates the current estimate of the latent channel states at each symbol. In addition, the channel is assumed to be Rayleigh fading under constant velocity following Jake's model \cite{tong2004paw}.
\subsection{Bayesian system model}
In this section we introduce the Bayesian system model that we consider for channel tracking and parameter inference.
In the context of this paper, online estimation refers to a frame by frame estimation procedure.
In this model, the channels in the relay network are treated as stochastic, where
we do not know \textit{a priori} the realized channel coefficient values.
We now present the model assumptions.
\vspace{0.3cm}
\subsubsection{Model Assumptions} \label{ModelAssumptions}
\begin{enumerate}
	\item[i.] \textit{Assume a wireless relay network with one mobile source node, transmitting symbols in frames of length $T$.}
	\item[ii.] \textit{The relays cannot transmit and receive on the same time slot and on the same frequency band. We thus consider a half duplex system model in which the data for a given frame are transmitted via a two step procedure. In the first step, the source node broadcasts a frame to all the relay nodes. In the second step, the relay nodes transmit the processed frame, termed the relay signals, to the destination node in \textbf{orthogonal fashion, ie. non-interfering channels}, see for example \cite{laneman2003dst}, \cite{laneman:2000}. In addition, we assume that all channels are independent with a coherence interval larger than the duration of the symbol.}
\item[iii.] \textit{We denote the mobile's angular Doppler frequency, relative to the $l$-th relay, by $\omega_m^{(l)}$ and it is assumed to be random unknown and constant throughout a frame .}
\item[iv.] \textit{We denote the $l$-th relay angular Doppler frequency by $\omega^{(l)}_r$, and it is assumed to be random unknown and constant throughout a frame and independent from the mobile, the other relays and the base station.}
\item[v.] \textit{We assume that all channels are flat fading, and our model is general enough to consider both scenarios of slow and fast fading.}
\item[vi.] \textit{The channels are parametrized under Jake's model \cite{jakes:1994}. The $l$-th relay channel is modeled as a two stage latent stochastic process, in which at time $n$ we denote the realization of the two channel stages by $h^{(l)}_n$ and $g^{(l)}_n$. The distribution of each stage of the channel at time $n$ is specified as}
\begin{equation}
\begin{split}
H^{(l)}_n & \sim \mathcal{CN}\left(0,\sigma^2_h\right) \; \; \; \text{Mobile} \Rightarrow \text{Relay channel}\\
G^{(l)}_n & \sim \mathcal{CN}\left(0,\sigma^2_g\right)\; \; \; \text{Relay} \Rightarrow \text{BS channel}.
\end{split}
\end{equation}
\textit{In addition we assume that the channels are temporally correlated and spatially i.i.d (between relays). This corresponds to the following model assumptions}
\begin{subequations}
\begin{align}
%\begin{split}
\mathbb{E}\left[ H^{(l)}_n G^{(j)}_k\right] &= 0 ,\; \forall n,k,l,j\\
\mathbb{E}\left[ H^{(l)}_n H^{(j)}_k\right] &= 0 , \; \forall n,k, l \neq j\\
\label{G_correlation}
\mathbb{E}\left[ G^{(l)}_n G^{(l)}_k\right] &= \sigma^2_h J_0\left(\omega^{(l)}_r \left|k-n\right|\right) ,\; \forall n,k, l\\
\label{H_correlation}
\mathbb{E}\left[ H^{(l)}_n H^{(l)}_k\right] &= \sigma^2_g J_0\left(\omega_m^{(l)} \left|k-n\right|\right) J_0\left(\omega^{(l)}_r  \left|k-n\right|\right) ,\; \forall n,k, l
%\end{split}
\end{align}
\end{subequations}
\textit{where, $J_0$ is the zeroth-order Bessel function of the first kind. For details, see Section II in \cite{patel2007channel}}.
%%%%%%%%%%%%%%%%%%
\item[vii.]{ \textit{The received signal at the $l$-th relay is a random variable given by}
\begin{align}
R^{(l)}_n = s_n H^{(l)}_n+ W^{(l)}_n, \; l \in \left\{1,\ldots,L\right\},
\end{align}
\textit{where at time $n$, $H^{(l)}_n$ is the channel coefficient between the transmitter
and the $l$-th relay, $s_n$ is the transmitted pilot symbol and $W^{(l)}_n$ is the unknown noise realization associated with
the relay receiver.}}

\item[viii.]{\textit{The received signals at the destination is a random variable given by}
\begin{align}
\label{ObservationEqn}
Y^{(l)}_n = f^{(l)}\left(R^{(l)}_n,n\right)G^{(l)}_n+V^{(l)}_n, \; l\in \left\{1,\ldots,L\right\},
\end{align}
\textit{
where at time $n$, $G^{(l)}_n$ is the channel coefficient between the $l$-th
relay and the receiver, $f^{(l)}\left(R^{(l)},n\right)$ 
\textit{is the memoryless relay processing function (with
possibly different functions at each of the relays) and $V^{(l)}_n$
is the noise realization associated with the relay receiver.}}
}
\item[ix.]{\textit{All received signals are corrupted by i.i.d. zero-mean additive white complex
Gaussian noise (AWGN). At the $l$-th relay the noise corresponding
to the $n$-th transmitted symbol is denoted by random variable
$W_n^{(l)} \sim \mathcal{CN} \left(0,\sigma_w^2\right)$. Then
at the receiver this is denoted by random variable $V_n^{(l)} \sim \mathcal{CN}\left(0,\sigma_v^2\right)$.
Therefore:
\begin{equation*}
\mathbb{E}\left[W_n^{(l)} \overline{W}_k^{(m)} \right]=
\mathbb{E}\left[V_n^{(l)} \overline{V}_k^{(m)} \right]=
\mathbb{E}\left[W_n^{(l)} \overline{V}_k^{(m)} \right]=0,
\end{equation*}
$\forall \left(n,k\right) \in \left\{1,\ldots,K\right\}, \forall \left(l,m\right) \in \left\{1,\ldots,L\right\}, n \neq k, l \neq m$
}}

\end{enumerate}
\vspace{0.3cm}
\subsubsection{Bayesian Model}
We begin by specifying the latent dynamic model for the channel coefficients used to approximate Jakes' model, as studied in \cite{Chang:1996} 
\begin{equation}
\label{Jakes}
\begin{split}
	H^{(l)}_n &= \alpha^{(l)} H^{(l)}_{n-1} + \sqrt{1- \left(\alpha^{(l)}\right)^2}\Upsilon^{(l)}_n\\
	G^{(l)}_n &= \beta^{(l)}  G^{(l)}_{n-1} + \sqrt{1- \left(\beta^{(l)}\right)^2} \Omega^{(l)}_n,
\end{split}	
\end{equation}
where $\Upsilon_n^{(l)}  \sim \mathcal{CN}\left(0,1\right)$, $\Omega_n^{(l)}  \sim \mathcal{CN}\left(0,1\right)$.

Next we formulate the Bayesian model, by first making precise the posterior parameters of interest in our Bayesian system model, $\bm{\alpha}, \bm{\beta}, \g_{1:T},\h_{1:T}$. To complete the information required to specify a Bayesian model we must present the remaining priors. The parameters for the channel coefficients are modeled as unknown \textit{a-priori} and represent the uncertainty in the system parameters $\sigma^2_h, \sigma^2_g, \omega^{(l)}_r,\omega^{(l)}_m$ given in eqs. (\ref{G_correlation}-\ref{H_correlation}). In particular since we consider $\sigma^2_h, \sigma^2_g, \omega^{(l)}_r,\omega^{(l)}_m$ to be unknown random variables, therefore we have that $\alpha^{(l)}$ and $\beta^{(l)}$ are also \textit{a priori} random variables given by
\begin{subequations}
\begin{align}
\beta^{(l)} &= \sigma^2_h J_0\left(\omega^{(l)}_r \left|k-n\right|\right) ,\; \forall n,k, l\\
\alpha^{(l)} &= \sigma^2_g J_0\left(\omega_m^{(l)} \left|k-n\right|\right) J_0\left(\omega^{(l)}_r  \left|k-n\right|\right) ,\; \forall n,k, l.
\end{align}
\end{subequations}
Hence, we specify priors on the parameters $\alpha^{(l)}$ and $\beta^{(l)}$ as follows,
\begin{equation}
\begin{split}
	\alpha^{(l)} \sim \text{Beta}\left(a,b\right), \\
	\beta^{(l)} \sim \text{Beta}\left(c,d\right) ,
\end{split}	
\end{equation}
where $\text{Beta}\left(x;a,b\right) \triangleq \frac{1}{\text{B}\left(a,b\right)}x^{a-1}\left(1-x\right)^{b-1}$, and $\text{B}(\cdot)$ is the beta function.\\
Furthermore, for simplicity, the same prior is used for all relay nodes. This prior choice is made to reflect the physical reality of the model.
More specifically, to insure a stationary model, a support of $\left[0,1 \right]$ is required for the prior. Secondly, it should be possible to insure that the majority of the prior mass is located close to the right boundary to insure realistic Doppler offsets scenarios.
 The choices of $a$, $b$ and $c$, $d$ reflect a realistic scenario of transmission in which the velocity of both transmitter and relay is practically achievable. 

\vspace{0.3cm}
\textbf{Remark 1} - The Bayesian model presented encompasses the case where the relays are both mobile or stationary. In case of stationary relays, the angular Doppler frequency of the $l$-the relay ,$\omega^{(l)}_r$, is set to $0$. 

\textbf{Remark 2} - We can now exploit directly the properties of the relay model proposed by considering the model assumptions which result in a non-linear state space model formulation for the system. The latent state processes corresponding to Jake's channel model at each stage of the relay system are given in eq. (\ref{Jakes}), these provide the state update equations for $\h_{1:T}$ and $\g_{1:T}$ as well as the unknown model parameters $\bm{\alpha}$ and $\bm{\beta}$. The non-linear observation equations, relating the transmitted data to the received data at the destination, after propagation through the relay network are given in eq. (\ref{ObservationEqn}). Note, the observation equation also introduces auxiliary variables to the state space model corresponding to the relay noise $\w_{1:T}$. Therefore the problem we address in this paper involves the following marginal posterior for model parameters and channel states 
\begin{equation}
\label{marg_posterior}
\begin{split}
%&p \left(\alpha^{(1:L)}, \beta^{(1:L)}, g_{1:T}^{(1:L)},h_{1:T}^{(1:L)}| y^{(l:L)}_{1:T} \right)
&p \left(\bm{\alpha}, \bm{\beta}, \g_{1:T},\h_{1:T}| \y_{1:T} \right)
\\
&\propto \prod_{l=1}^L \left[\prod_{n=1}^T p \left(y^{(l)}_{n}|\alpha^{(l)}, \beta^{(l)},g_{n}^{(l)},h_{n}^{(l)} \right)
														p \left(h_{n}^{(l)}|h_{n-1}^{(l)} \right)
														p \left(g_{n}^{(l)}|g_{n-1}^{(l)} \right)\right]
														p \left(g_{1}^{(l)} \right)
														p \left(h_{1}^{(l)} \right)
														p \left(\alpha^{(l)} \right)
														p \left(\beta^{(l)} \right),
\end{split}														
\end{equation}
where 
$\alpha^{(1:L)}, \beta^{(1:L)}, g_{1:T}^{(1:L)},h_{1:T}^{(1:L)}, y^{(l:L)}_{1:T} \triangleq \bm{\alpha}, \bm{\beta}, \g_{1:T},\h_{1:T}, \y_{1:T}$.
In addition it is worth noting that this decomposition of the posterior distribution utilizes directly the relay model structure. This posterior model that we aim to estimate is very high dimensional with $L(3T + 2)$ parameters and not tractable for standard Bayesian estimation procedures. For this reason we require advanced computational tools such as PMCMC that will be presented in Section \ref{ChannelTrackingSamplingFramework}. These techniques are specifically designed to efficiently allow one one to work with such statistical models in relay networks with multiple relays and multiple hops even in high dimensions.
 %%%%%%%%%%%%%%%%%%%%%%%%%%%%%%5
\vspace{0.1cm}
\begin{corollary}{\textit{
As a consequence of the model assumptions, the posterior distribution factorises according to the following independence structure
\begin{equation}
\begin{split}
p \left(\bm{\alpha}, \bm{\beta}, \g_{1:T},\h_{1:T}| \y_{1:T} \right)
=\prod_{l=1}^L p \left(\alpha^{(l)}, \beta^{(l)},g_{1:T}^{(l)},h_{1:T}^{(l)}|y^{(l)}_{1:T} \right),
\end{split}														
\end{equation}
with respect to the number of parallel relay transmission paths.}}\\
%%%%%%%%%%%%%%%%%%%%%%%%%%%%%%%%%%%%%%%
%%%%%%%%%%%%%%%%%%%%%%%%%%%%%%%%%%5
\end{corollary}
%%%%%%%%%%%%%%%%%%%%%%%

\textbf{Remark 3} - Since the posterior factorises to produce independence between relay transmission paths, this enables us to exploit this structure in the design of the estimation framework. In particular, the approach that we take which involves PMCMC now admits a natural block factorisation structure in which the particle filters per block may be run independently. The details relating to this remark will be presented in Section \ref{ChannelTrackingSamplingFramework}. This means that we are able to estimate the proposal distribution, in the PMCMC algorithm for the channels trajectories, via parallel independent particle filters. As a result of this, the variance of the incremental important weights used in the estimation of the proposal can be reduced leading to an increased PMCMC acceptance probability. \textbf{Therefore, under the model and estimation procedure proposed, increasing the number of relays will not degrade the solution.} These properties of the system model are well accepted, see \cite{laneman:2000}. However, the estimation procedures we present are general and can be applied also to solutions with spatial correlation, in which this decomposition no longer applies.
\vspace{0.1cm}

%%%%%%%%%%%%%%%%%%%%%%%%%%%%%%5
\begin{corollary}
{\textit{
For relay network topologies with multiple hops, $K$, per parallel transmission path $l$, the posterior distribution factorises according to the following independent structure
\begin{equation}
\begin{split}
p \left(\bm{\theta}, \bm{\Lambda}_{1:T}| \y_{1:T} \right)
= \prod_{l=1}^L p \left(\bm{\theta}^{(l)},\bm{\Lambda}_{1:T}^{(l)}|y^{(l)}_{1:T} \right),
\end{split}														
\end{equation}
where $\bm{\theta}^{(l)} =\left[\theta^{(l,0)},\ldots, \theta^{(l,K)} \right]$ corresponds to the unknown static parameters for the $l$-th path; and 
$\bm{\Lambda}_{n}^{(l)} =\left[ \Lambda_{n}^{(l,0)},\ldots, \Lambda_{n}^{(l,K)} \right]$ corresponds to the channel gain at the $n$-th epoch of the $l$-th relay path.\\
%%%%%%%%%%%%%%%%%%%%%%%%%%%%%%%%%%%%%%%
%%%%%%%%%%%%%%%%%%%%%%%%%%%%%%%%%%5
}}
\end{corollary}
%%%%%%%%%%%%%%%%%%%%%%%
\textbf{Remark 4} - From this we see how the methodology scales as a function of the number of relay hops $K$. We note that as $K$ increases one should be careful to ensure the proposal for the $K$ channels in the $l$-th relay link for the particle filtering aspect of the PMCMC algorithm is designed to approximate the optimal importance distribution for each time $n$. The reason for this is that as the dimension of the state at epoch $n$ for a given particle filter block increases, one must always be careful to reduce the variance of the incremental importance sampling weights, therefore improving the acceptance rate of the PMCMC algorithm. \textbf{Therefore, under the model and estimation procedure proposed, increasing the number of hops will increase the variance of the solution obtained.} The details regarding this remark are presented below.

\begin{figure}[b]
    \centering
        \epsfysize=10cm
        \epsfxsize=15cm
        \epsffile{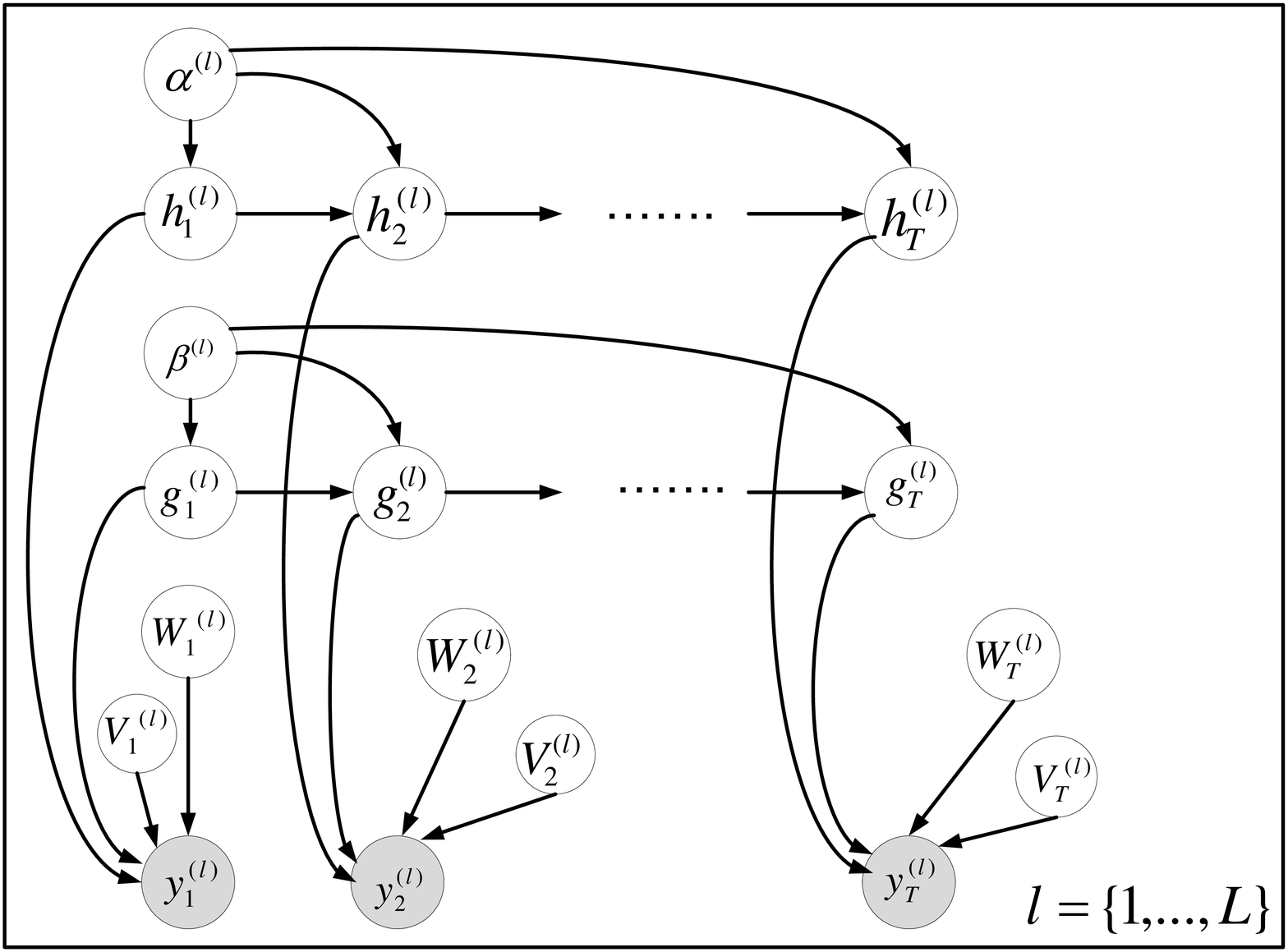}
        \caption{Graphical model of the relay system with a dual hop}
    \label{fig:GraphicalModel}
\end{figure}

The relay system statistical model proposed can be summarised by the graphical model structure presented in Fig. \ref{fig:GraphicalModel}.
%%%%%%%%%%%%%%%%%%%%%%%%%%%%%%%
%%%%%%%%%%%%%%%%%%%%%%%%%%%%%%%
\section{Channel Tracking Sampling Framework} \label{ChannelTrackingSamplingFramework}
In this section we first discuss the challenges associated with estimating the model and performing channel tracking under the Bayesian relay model framework developed in Section \ref{ModelSection}. We begin by outlining a common approach to tackle the problem of joint static parameter and latent dynamic process estimation. 
We discuss the complications that arise when using this common approach to solve the channel tracking problem. We then present a novel sampling algorithm to overcome these difficulties, based on Adaptive MCMC and Particle MCMC, which we term the Adaptive PMCMC (AdPMCMC) algorithm. 

The channel tracking problem involves developing an efficient algorithm to estimate the unknown vector of parameters $\bm{\alpha},\bm{\beta}$ and the vector of latent process channels states $\H_{1:T},\G_{1:T}$. To achieve this we consider the augmented Bayesian posterior 
$p \left(\bm{\alpha}, \bm{\beta}, \g_{1:T},\h_{1:T},\w_{1:T}|\y_{1:T} \right)$, containing auxiliary variables $\W_{1:T}$ which we marginalise out numerically in our sampling algorithm to obtain the posterior corresponding to (\ref{marg_posterior}). 
Under this posterior distribution, we can find a solution to the channel tracking problem which involves obtaining point estimates such as the Maximum a-posteriori MAP (posterior mode), the Minimum Mean Square Error MMSE (posterior mean) and posterior credible intervals for $\left\{\bm{\alpha}, \bm{\beta}, \G_{1:T},\H_{1:T}\right\}$, given $\y_{1:T}$. In this paper we refer to efficiency of a sampling algorithm as related to the mixing rate of the Markov chain. That is how rapidly the Markov chain can reach the stationary regime of the posterior from an arbitrary intialisation. In particular the more efficient the mixing rate of an algorithm, for a given computational complexity, the more accurate the estimated posterior quantities will be, therefore with more accurate channel estimates, we can then improve resulting estimation challenges such as synchronisation, power allocation and percoding.

\subsection{Standard MCMC-Gibbs sampler approach}
There are several sampling approaches that can be considered for state space models involving unknown states and parameters.
Details and discussion of standard algorithms in this context can be found in \cite{bar2001estimation}, \cite{durbin2001time}, \cite{silva2009particle}, \cite{carter1994gibbs} and the references therein. 
Such approaches in the context of state space models are typically inefficient. This is because they are susceptible to inefficiencies arising from the very high dimension of the problem, especially when correlation is present in the posterior. Therefore, though technically valid, in many practical settings this precludes the use of naive sampling strategies from a practical computational cost. Typically these approaches involve splitting the high dimensional posterior distribution into subblocks of parameters and then running a blockwise Metropolis-Hastings (MH) within Gibbs sampling framework. One such approach that we consider to compare to our AdPMCMC methodology is based on a basic Markov chain Monte Carlo algorithm involving a Gibbs sampling framework.
Basically this involves a sampling framework in which the posterior denoted $p \left(\bm{\alpha},\bm{\beta}, \g_{1:T},\h_{1:T},\w_{1:T}| \y_{1:T} \right)$ is sampled by splitting the vector of latent states into $k$ sub-blocks of length $\tau$, where $k \tau = T$, and each iteration of the Markov chain updates each sub-block of the states in either a deterministic or random scan until a Markov chain of length $J$ is obtained. This is presented in Algorithm \ref{Standard_MHinGibbs}, and is just one of many possible block structures that could be used. The full conditional posterior distributions are typically sampled from via a MH-within-Gibbs sampling framework.

The simplest approach is to sample the univariate full conditional distributions, ie. $k=T$ blocks of size $\tau = 1$. 
However, this sampling scheme will typically result in very slow mixing of the Markov chain around the support of the posterior, making this naive MCMC algorithm computationally impractical. This is
especially problematic in high dimensional target posterior distributions, leading to poor channel estimates with high variance. It is well known that to avoid this slow mixing Markov chain setting, one must sample from larger blocks of parameters. However, the design of an optimal proposal distribution for large blocks of parameters is very complicated. 
The efficiency of such a naive block Gibbs sampling algorithm is dependent both on the choice of blocking of the posterior parameters, the size of the blocks updated at each stage and the sampling mechanism for each block. In general it is a significant challenge in practice to design algorithms which are efficient in such block Gibbs settings when correlation is present between the parameters of the posterior distribution, as will occur in non-linear state space setting of channel tracking.
As a result for moderate sized values of $\tau$ the simple MH-within-Gibbs framework can be poorly mixing, due to low acceptance probabilities, even when carefully design proposals are implemented. This leads to requirements for very long Markov chain lengths which is not practical. We overcome these well known problems of naive MCMC sampling algorithms by developing a novel version of the PMCMC methodology \cite{andrieu-particle} utilizing adaptive MCMC and SMC algorithms in a nonstandard manner as proposal structures, explained below. In section \ref{AdMCMCDetails} we explain in detail the properties and specification for this non-standard algorithm. The methodological innovation we present in this paper is to combine an Adaptive MCMC algorithm within the PMCMC framework with a Rao-Blackwellised SIR particle filter allowing us to update $\left(\bm{\alpha}, \bm{\beta}, \g_{1:T},\h_{1:T},\w_{1:T}\right)$ in a single efficient iteration of the Markov chain. 

The Markov chain state vector, for the relay model in Section \ref{ModelSection}, is in a very high dimension of $L(3T + 2)$ parameters. It is therefore critical to any MCMC mechanism to attempt to approximate the optimal proposal distribution.
The AdPMCMC sampler we develop achieves this by approximating the optimal proposal, through a combination of an Adaptive MCMC and Rao-Blackwellised SIR particle filter. This improves the MCMC algorithm significantly. 

\subsection{Advanced Adaptive MCMC within Rao-Blackwellised Particle MCMC (AdPMCMC).}
\label{AdMCMCDetails}
The aim of this section is to present a novel methodology to perform sampling form the posterior distribution given in eq. (\ref{marg_posterior}). \\
%The simulation methodology extended here is based on \cite{andrieu-particle} and represents a state of the art sampling framework for such problems. \\
The PMCMC approach we develop works by approximating the marginal Metropolis Hastings algorithm. It therefore updates at each iteration of the Markov chain, the joint channel and parameter vector 
$\left[\bm{\alpha}, \bm{\beta}, \h_{1:T},\g_{1:T},\w_{1:T}\right]$ in an efficient manner. To achieve this the PMCMC algorithm involves approximation of the optimal proposal distribution for the latent states $p(\h_{1:T},\g_{1:T},\w_{1:T}|\bm{\alpha}, \bm{\beta},\y_{1:T})$ with a particle filter estimate on the path space, which can be easily sampled from. In addition, the marginal likelihood $p(\y_{1:T}|\bm{\alpha}, \bm{\beta})$ which is used in the evaluation of the acceptance probability can not be obtained analytically. This is due to the fact that marginalization of the joint likelihood over the path space involves the following integration
\begin{equation*}
\begin{split}
p(\y_{1:T}|\bm{\alpha}, \bm{\beta}) &= \prod_{t=0}^{T-1}p(\y_{t+1}|\y_{1:t},\bm{\alpha}, \bm{\beta})\\
&= \int \ldots \int \left[\prod_{t=0}^{T-1}  p(\y_{t+1}|\h_{t+1},\g_{t+1},\w_{t+1},\bm{\alpha}, \bm{\beta}) p(\h_{t+1},\g_{t+1},\w_{t+1}|\y_{1:t}, \bm{\alpha}, \bm{\beta})\right]d\h_{1:T} d\g_{1:T} d\w_{1:T}
\end{split}
\end{equation*}
which can not be performed analytically. Note we define $y_{1:0} = \emptyset$. It can however also be estimated efficiently using the same particle filter used in the proposal estimate. Remarkably the key result of \cite{andrieu-particle} was to prove that for the PMCMC algorithm utilizing these two particle filter approximations, no matter how many particles are utilised in SMC algorithm for each approximation, the stationary distribution of the PMCMC Markov chain is unbiased. In particular for our model, the stationary distribution of our PMCMC algorithm still remains the target posterior distribution in (\ref{marg_posterior}).\\
%For details of the proof of this remark, see \cite{andrieu-particle}.\\
The Particle MCMC proposal distribution to move from a stat at iteration $j$ to a new state at iteration $(j+1)$ is split into two components. 
\begin{align}
\label{proposal_lernel}
\begin{split}
&q\left(\left[\bm{\alpha}, \bm{\beta}, \g_{1:T},\h_{1:T},\w_{1:T}\right](j);
\left[\bm{\alpha}, \bm{\beta}, \g_{1:T},\h_{1:T},\w_{1:T}\right](j+1)
\right)\\
&=q\left(\left[\bm{\alpha}, \bm{\beta}\right](j);\left[\bm{\alpha}, \bm{\beta}\right](j+1)\right)
p\left(\left[\h_{1:T},\g_{1:T},\w_{1:T}\right](j+1) |  \y_{1:T},
\left[\bm{\alpha},\bm{\beta}\right](j+1)\right).
\end{split}
\end{align}
The first involves a proposal kernel which is constructed via an adaptive MH scheme and this will be
used to sample the static parameters $\bm{\alpha}, \bm{\beta}$. The second component of the proposal kernel involves the sampling of a trajectory for the latent channels and relay noise, $\g_{1:T},\h_{1:T},\w_{1:T}$. The resulting PMCMC approach is presented in Algorithm \ref{Generic_PMCMC}. The remainder of this section will first detail the Adaptive MCMC component of the PMCMC proposal mechanism, followed by the filtering component of the proposal. This section is completed with the details of the marginal MH acceptance probability for the AdPMCMC algorithm.

The introduction of an adaptive MCMC proposal kernel into the Particle MCMC setting allows the Markov chain proposal distribution to adaptively learn the regions in which the marginal
posterior distribution for the static model parameters has most mass. As such the probability of acceptance under such an adaptive proposal will be significantly improved over time. 
Then for the latent channels processes we develop the non-standard PMCMC proposal kernel constructed via a Sequential Monte Carlo algorithm which will be based on a Rao-Blackwellised SIR filter \cite{doucet2000rao}. The Rao-Blackellisation is performed via a conditional Kalman filter structure. The conditioning is specifically chosen to allow one to work with arbitrary numbers of relay hops and also more importantly an arbitrary non-linear relay function. In particular, this involves a particle filter for $\h_{1:T},\w_{1:T}$ and a conditional Kalman filter proposal for $\g_{1:T}$.

In the state space setting, the PMCMC algorithm used to sample from a target distribution (\ref{marg_posterior}) proceeds by mimicking the marginal MH algorithm \cite{andrieu-particle} and \cite{andrieu2009pseudo} in which the acceptance probability, denoted by A, going from state at iteration $j$ of the Markov chain to iteration $(j+1)$, is given by
\begin{align}
\label{AccptProb}
\begin{split}
&A \left(\left[\bm{\alpha},\bm{\beta},\h_{1:T},\g_{1:T},\w_{1:T}\right](j);
%%%%%%%%%%%%
\left[\bm{\alpha},\bm{\beta},\h_{1:T},\g_{1:T},\w_{1:T}\right](j+1)
\right) = \\
&\min \left( 1,
\frac{p\left(\left[\bm{\alpha}, \bm{\beta}\right](j+1)  |\y_{1:T}\right)
q\left(\left[\bm{\alpha}, \bm{\beta}\right](j+1);\left[\bm{\alpha}, \bm{\beta}\right](j)\right)}
{p\left(\left[\bm{\alpha}, \bm{\beta}\right](j)|\y_{1:T}\right)
q\left(\left[\bm{\alpha}, \bm{\beta}\right](j);\left[\bm{\alpha}, \bm{\beta}\right](j+1)\right)}
\right).
\end{split}
\end{align}
Clearly, achieving this requires one to use a very particular
structure for the proposal kernel in the MCMC algorithm, given in eq. (\ref{proposal_lernel}). In particular after substitution of this proposal into the standard MH acceptance probability, we obtain the marginalised form given by (\ref{AccptProb}).

\textbf{The critical idea in formulating the Particle MCMC algorithm is that the proposal distribution for
$p\left(\left[\h_{1:T},\g_{1:T},\w_{1:T}\right](j+1)
|\y_{1:T},\left[\bm{\alpha}, \bm{\beta}\right](j+1) \right)$ can be sampled from approximately via a Sequential Monte Carlo algorithm (otherwise known as a
particle filter)}. Sequential Monte Carlo, \cite{doucet2009tutorial}, \cite{del2004feynman}, \cite{doucet2000sequential} refers to a class of algorithms which have become popular due to their algorithmic and theoretical properties, especially in filtering problems in which a non-linear or non-Gaussian state space model is considered.

In \cite{andrieu-particle}, the PMCMC sampler has been shown to have several theoretical convergence properties. In particular, \cite{del2004feynman} derives convergence in the
empirical law of the particles to the true filtering distribution at each iteration is bounded as a linear function of time $t$ and the number of particles $N$. In addition a Central Limit Theorem can be obtained,
\begin{align*}
&\parallel \mathcal{L}\text{aw}\left(	\left[\h_t,\g_t,\w_t\right](j,i) \right) -
p\left(\h_t,\g_t,\w_t|\y_{1:t},\left[\bm{\alpha}, \bm{\beta}\right](j)\right)
\parallel \leq C\frac{t}{N}, \\
&\left(\widehat{p}\left(\y_{t}|\bm{\alpha}, \bm{\beta}\right)-p\left(\y_{t}|\bm{\alpha}, \bm{\beta}\right) \right) \rightarrow
\mathcal{N}\left(0,\sigma_t^2\right),
\end{align*}
where $p\left(\y_{t}|\bm{\alpha}, \bm{\beta}\right) $ is the normalising constant of the filtering distribution and $\sigma_t^2 \leq D\frac{t}{N}$. These results are
important as they demonstrate that the complexity of the problem
only scales linearly with dimension.

The other innovation we introduce to the proposal mechanism of the PMCMC algorithm involves developing a non-trivial adaptive MCMC approach for the static parameters proposal ($q\left(\left[\bm{\alpha}, \bm{\beta}\right](j);\left[\bm{\alpha}, \bm{\beta}\right](j+1)\right)$) in (\ref{proposal_lernel}).
There are several classes of adaptive MCMC algorithms, see \cite{roberts2009examples}. The distinguishing feature of adaptive MCMC
algorithms, compared to standard MCMC, is that they utilise a combination of time or state inhomogeneous proposal kernels. Several recent papers proposed theoretical conditions that our approach satisfies, ensuring ergodicity of our adaptive algorithms, see \cite{silva2009particle}, \cite{atchade2005adaptive}, \cite{haario2001adaptive}, \cite{andrieu2006ergodicity} and \cite{andrieu2006efficiency}.

In \cite{roberts2009examples} ergodicity of adaptive MCMC is proved under conditions known as \textit{Diminishing Adaptation}
and \textit{Bounded Convergence}. As in \cite{roberts2009examples} we assume that each fixed kernel in the sequence
$Q_{\gamma}$ has stationary distribution $P\left(\cdot\right)$ which corresponds to the marginal posterior of the static parameters.
Define the convergence time for kernel $Q_{\gamma}$ when starting
from state $\bm{\alpha}, \bm{\beta}$ as $M_{\epsilon}\left(\bm{\alpha}, \bm{\beta},\gamma\right) =
\text{inf}\{s \geq 1 : \|Q^s_{\gamma}\left(\bm{\alpha}, \bm{\beta};\cdot\right) -
P\left(\cdot\right)\| \leq \epsilon $. Under these assumptions,
they derive the sufficient conditions;
\begin{itemize}
\item{ \textbf{Diminishing Adaptation:} $\text{lim}_{n\to\infty}\text{sup}_{\bm{\alpha}, \bm{\beta} \in E}\|Q_{\Gamma_{s+1}}\left(\bm{\alpha}, \bm{\beta},\cdot\right) - Q_{\Gamma_{s}}\left(\bm{\alpha}, \bm{\beta},\cdot\right)\| = 0$ in probability. Note, $\Gamma_s$ are random indices.}
\item{ \textbf{Bounded Convergence:} $\{M_{\epsilon}\left(\left[\bm{\alpha}, \bm{\beta}\right](j),\Gamma_j\right)\}^\infty_{j=0}$ is bounded in probability, $\epsilon > 0.$}
\end{itemize}
which guarantee asymptotic convergence in two senses,
\begin{itemize}
\item{\textbf{Asymptotic convergence:} $\text{lim}_{j\to\infty}\|\mathcal{L}\text{aw}\left(\left[\bm{\alpha}, \bm{\beta}\right]	(j)\right)-P\left(\bm{\alpha}, \bm{\beta}\right)\|=0$}
\item{\textbf{Weak Law of Large Numbers}: $\text{lim}_{j\to\infty}\frac{1}{j}\sum^{j}_{i=1}\phi \left(\left[\bm{\alpha}, \bm{\beta}\right](i)\right)=\int \phi(\bm{\alpha}, \bm{\beta})P(d\bm{\alpha}, d\bm{\beta}) $ for all bounded $\phi : E \to R$.}
\end{itemize}

\textbf{Algorithmic Choices and Specifications for the AdPMCMC Algorithm}\\
We present the specific details of the adaptive MH within Particle MCMC algorithm used to sample from the
posterior on the path space of our latent factors and state space model parameters. 
\subsubsection{Adaptive MCMC for static parameters $\bm{\alpha}$, $\bm{\beta}$}
here we detail the specifics of step 3 in Algorithm \ref{Generic_PMCMC}, which involves specification of the proposal distribution in the Particle MCMC algorithm (see eq. (\ref{proposal_lernel})).
The static parameters are updated via an adaptive MH proposal comprised of a mixture of Gaussians. One of the mixture components has a covariance structure which is adaptively learnt on-line. The mixture proposal distribution for parameters $\left[\bm{\alpha}, \bm{\beta}\right]$ at iteration $j$ of the Markov chain is given by,
\begin{equation}
\label{AMCMC_random_walk}
\begin{split}
q\left(\left[\bm{\alpha}, \bm{\beta}\right](j);\left[\bm{\alpha}, \bm{\beta}\right](j+1)\right)&=
w_1 N\left(\left[\bm{\alpha}, \bm{\beta}\right](j+1);\left[\bm{\alpha}, \bm{\beta}\right](j),\frac{\left(2.38\right)^2}{d}\Sigma_j\right)\\
&+ \left(1-w_1\right)
N\left(\left[\bm{\alpha}, \bm{\beta}\right](j+1);\left[\bm{\alpha}, \bm{\beta}\right](j),\frac{\left(0.1\right)^2}{d}I_{2L,2L}\right).
\end{split}
\end{equation}
Here, $\Sigma_j$ is the current empirical estimate of the covariance between the parameters of $\bm{\alpha}, \bm{\beta}$ estimated using
samples from the PMCMC chain up to time $j$, and $w_1$ is a mixture proposals weight which we set according to the recommendation of \cite{roberts2009examples}. The theoretical motivation for the choices of scale factors $2.38$, $0.1$ and dimension $d$ are all provided in \cite{roberts2009examples} and are based on optimality conditions presented in \cite{roberts2001optimal}.
The description is depicted in Algorithm \ref{adaptice_MCMC}.
%%%%%%%%%%%%%%%%%5
\subsubsection{Rao-Blackwellised SIR filter specifications}
the proposal kernel for the unknown channels $\g_{1:T},\h_{1:T}$ and auxiliary variables $\w_{1:T}$, denoted  by
$p\left(\g_{1:T},\h_{1:T}, \w_{1:T}|\y_{1:T},\left[\bm{\alpha}, \bm{\beta}\right](j)\right)$,  
involves the SIR particle filter in which Rao-Blackwellisation is achieved via a Kalman filter. As such, it is an approximation to the optimal proposal.
The following decomposition is used
\begin{equation}
p\left(\g_{t},\h_{t}, \w_{t}|\y_{1:t},\left[\bm{\alpha}, \bm{\beta}\right](j)\right)=
\underbrace{p\left(\g_{t}|\h_{t}, \w_{t},\y_{1:t},\left[\bm{\alpha}, \bm{\beta}\right](j)\right)}_{\text{Kalman filter}}
\times 
\underbrace{ p\left(\h_{t}, \w_{t}|\y_{1:t},\left[\bm{\alpha}, \bm{\beta}\right](j)\right)}_{\text{Particle filter}},
\end{equation}
where for each particle in the SIR filter for $\h_t,\w_t$, there is a corresponding Kalman filter for $\g_t$. This is presented in Algorithm \ref{SIR_filter}.

%%%%%%%%%%%%%%%%%%%%%%%%%%%%%%%%%%%%%%%%%%%%%%%%%%%%%%%%%%%%%%%%%%%%%%%%%%%%%%%%
\section{Cram\'er-Rao Lower Bound for the Path Space Proposal in PMCMC} \label{CRLB}
%%%%%%%%%%%%%%%%%%%%%%%%%%%%%%%%%%%%%%%%%%%%%%%%%%%
%%%%%%%%%%%%%%%%%%%%%%%%%%%%%%%%%%%%%%%%%%%%%%%%%%%%
%\subsection{Synthetic Data - Mean Square Error Analysis}
%%%%%%%%%%%%%%%%%%%%%%%%%%%%%%%%%%%%%%%%%%%%%%%%%%%%
In this section we study the Mean Square Error for estimation of the relay channel estimations after integrating out the uncertainty in the static parameters $\bm{\alpha},\bm{\beta}$, for a range of SMC particle counts N. This will provide for us an understanding of the accuracy of our methodology AdPMCMC in estimating the true underlying process for a given signal to noise ratio. We also derive a recursive expression for the Bayesian Cram\'er-Rao Lower Bound (BCRLB) as a lower bound comparison. \textit{We demonstrate how the BCRLB can be trivially estimated at no additional computational cost in our model framework, recursively for each time step $t$, via the AdPMCMC algorithm and a modified recursion from \cite{tichavsky1998}.} We derive analytically these results for any unbiased estimate of the marginal latent processes, $\g_{1:T},\h_{1:T}, \w_{1:T}$,  conditional on a realisation of the static model parameters, $\left[\bm{\alpha}, \bm{\beta}\right]\left(j\right)$.
We then show how this can be calculated recursively on the path space at each iteration of the PMCMC algorithm for each realized data set, allowing us to numerically evaluate
\begin{align}
\begin{split}
\label{MC_CRLB}
&\int\cdots\int  \left\{\left[\X_{1:T}-\widehat{\X}_{1:T}\right]\left[\X_{1:T}-\widehat{\X}_{1:T}\right]^H\right\} 
p\left(\x_{1:T},\y_{1:T}, \bm{\alpha},\bm{\beta} \right) d\x_{1:T} d \y_{1:T}   d \bm{\alpha} d \bm{\beta} \\
%%%%%%%%%5
&=\int\cdots\int \left\{\left[\X_{1:T}-\widehat{\X}_{1:T}\right]\left[\X_{1:T}-\widehat{\X}_{1:T}\right]^H\right\} 
p\left(\x_{1:T},\y_{1:T}| \bm{\alpha},\bm{\beta} \right) 
p\left(\bm{\alpha},\bm{\beta} \right)
d\x_{1:T} d \y_{1:T}   d \bm{\alpha} d \bm{\beta} \\
%%%%%%%%%%%%
&=\int\cdots\int \exE_{p(\x_{1:T},\y_{1:T}|\bm{\alpha},\bm{\beta})} \left\{\left[\X_{1:T}-\widehat{\X}_{1:T}\right]\left[\X_{1:T}-\widehat{\X}_{1:T}\right]^H|\bm{\alpha},\bm{ \beta}\right\}p\left(\bm{\alpha},\bm{\beta}\right) d \bm{\alpha} d \bm{\beta},
\end{split}
\end{align}
where $\x_{1:T} \triangleq \left[\h_{1:T}, \g_{1:T}, \w_{1:T}\right]$. Note, the Monte Carlo integration involved in the estimation of eq. (\ref{MC_CRLB}), does not require any additional computational complexity, as it is evaluated online at each iteration of the PMCMC algorithm for each data set generated. This BCRLB provides a lower bound on the MSE matrix for the path space parameters which correspond in our model to the estimation of the relay channel models.

We denote the CRLB by $\left[F_{1:T}\left(\g_{1:T},\h_{1:T}, \w_{1:T}\right)\right](j)$ and $\left[F_{t}\left(\g_{t},\h_{t}, \w_{t}\right)\right](j)$ will denote the Fisher Information Matrix (FIM) for symbol $t$ of the frame in the path space, conditional on the proposed static parameters at iteration $j$ of the PMCMC algorithm.

\noindent \textbf{Cram\'er-Rao Lower Bound for the Path Space Proposal in PMCMC} \label{CRLB}\\
In the Bayesian context we do not require that the estimator of interest, in our case $\X_{1:T}^{MMSE}$, be unbiased. However we do require that the model is specified such that the following two conditions hold. \\
\noindent \textbf{Condition 1:} for static state space model parameters for each relay link $l$ which are within some interval $\alpha^{(l)} \in [0,1]$ and $\beta^{(l)} \in [0,1]$, the prior models $p(\alpha^{(l)})$ and $p(\beta^{(l)})$ satisfy the conditions $\lim_{\alpha^{(l)}\rightarrow 0}p(\alpha^{(l)}) \rightarrow 0$; $\lim_{\beta^{(l)}\rightarrow 0}p(\beta^{(l)}) \rightarrow 0$; $\lim_{\alpha^{(l)}\rightarrow 1}p(\alpha^{(l)}) \rightarrow 0$ and $\lim_{\beta^{(l)}\rightarrow 1}p(\beta^{(l)}) \rightarrow 0$\\
\noindent \textbf{Condition 2:} The following smoothness properties of the likelihood hold: 
$$\int \frac{\partial p \left( \y_{1:T}|\bm{\alpha},\bm{\beta}  \right)}{\partial (\bm{\alpha},\bm{\beta})} d \y_{1:T} = 0.$$
Under these conditions we may then utilize the results of \cite{tichavsky1998} in which recursive expressions for the BCRLB are derived for general non-linear state space models. In particular we consider the recursion in time $t$, for the Bayesian equivalent of the Fisher information matrix on the estimate of $\X_t^{MMSE}$, given in eq. (21) of the paper. \textit{We modify these results to integrate out the posterior uncertainty in the joint estimation of the static parameters $\bm{\alpha},\bm{\beta}$ parameterizing the state space models.} In particular we derive results which can perform this marginalization numerically utilizing the existing AdPMCMC framework.

The BCRLB provides a lower bound on the MSE matrix for estimation of the path space parameters which correspond in our model to the estimation of the latent process states $\X_{1:T}$. We denote the Fisher Information Matrix (FIM), used in the CRLB, on the path space by $[F_{1:T}\left(\x_{1:T}\right)]\left(j\right)$ and marginally by $[F_{t}\left(\x_{t}\right)]\left(j\right)$ for time $t$ in the path space, conditional on the proposed static parameters at iteration $j$ of the PMCMC algorithm. Here we derive an analytic recursive expression for this quantity based on \cite{tichavsky1998}. In some cases we can get analytic solutions and in others, we will resort to AdPMCMC based online approximations with a novel estimation based on the particle filter proposal distribution of our AdPMCMC algorithm. 

Conditional on the previous Markov chain state $\left[\bm{\alpha},\bm{\beta}, \X_{1:T}\right]\left(j-1\right)$ and the new sampled Markov chain proposal for the static parameters at iteration $j$, $\left[\bm{\alpha},\bm{\beta}\right]\left(j\right) $, we obtain the following modified recursive expression for the FIM based on eq. (21) in \cite{tichavsky1998}:
\begin{equation}
\left[J_t(\widehat{\X}_t)\right](j) = \left[D_{t-1}^{22}(\widehat{\x}_t)\right](j) -\left[D_{t-1}^{21}(\widehat{\x}_t)\right](j) \left(\left[J_{t-1}(\widehat{\x}_t)\right](j)+\left[D_{t-1}^{11}(\widehat{\x}_t)\right](j)\right)^{-1}\left[D_{t-1}^{12}(\widehat{\x}_t)\right](j),
\end{equation}
where we obtain the following matrix decompositions of our system model, via Eqs. (34-36) of \cite{tichavsky1998} under the model assumptions of additive Gaussian process and observation noise: 
\begin{equation}
\begin{split}
\left[J_1(\widehat{\x}_1)\right](j) &= -\mathbb{E} \left[\bold \nabla_{\x_{1}}\left\{\bold \nabla_{\x_{1}} \log p\left(\x_1\right)
\right\}^T \right];\\
\left[D_{t-1}^{11}\right](j)  &= -\mathbb{E} \left[\bold \nabla_{\x_{t-1}}\left\{\bold \nabla_{\x_{t-1}} \log p\left(\x_t|\x_{t-1}\right)
\right\}^T \right] = \mathbb{E} \left\{ \left[\nabla_{\x_{t-1}}f\left(\x_{t-1};\bm{\alpha},\bm{\beta}\right)\right]Q_{t-1}^{-1}\left[\nabla_{\x_{t-1}}f\left(\x_{t-1};\bm{\alpha},\bm{\beta}\right)\right]^T\right\};\\
\left[D_{t-1}^{12}\right](j) &= \left[D_{t-1}^{21}\right](j) = -\mathbb{E} \left[\bold \nabla_{\x_{t}}\left\{\bold \nabla_{\x_{t-1}} \log p\left(\x_t|\x_{t-1}\right)\right\}^T \right] = -\mathbb{E} \left[\bold \nabla_{\x_{t-1}} f\left(\x_{t-1};\bm{\alpha},\bm{\beta}\right)\right]Q_{t-1}^{-1};\\
\left[D_{t-1}^{22}\right](j)  &= -\mathbb{E} \left[\bold \nabla_{\x_{t}}\left\{\bold \nabla_{\x_{t}} \log p\left(\x_t|\x_{t-1}\right)
\right\}^T \right] + -\mathbb{E} \left[\bold \nabla_{\x_{t}}\left\{\bold \nabla_{\x_{t}} \log p\left(\y_t|\x_{t}\right)
\right\}^T \right]\\
&= Q_{t-1}^{-1} + \mathbb{E}\left\{\left[\bold \nabla_{\x_{t}} h\left(\x_{t};\bm{\alpha},\bm{\beta}\right) \right]R_{t}^{-1}\left[\bold \nabla_{\x_{t}} h\left(\x_{t};\bm{\alpha},\bm{\beta}\right) \right]^T\right\}
\end{split}
\end{equation}
where $f\left(\x_{t-1};\bm{\alpha},\bm{\beta}\right)$ is the state model with process noise covariance $Q_t$ and $h\left(\x_{t};\bm{\alpha},\bm{\beta}\right)$ is the observation model with observation noise covariance $R_t$. Next we derive these quantities for each model conditional on the previous Markov chain state $\left[\bm{\alpha}, \bm{\beta}, \g_{1:T},\h_{1:T}, \w_{1:T}\right]\left(j-1\right) \triangleq 
\left[\bm{\alpha}, \bm{\beta}, \x_{1:T}\right]\left(j-1\right) $ and the new sampled Markov chain proposal for the static parameters at iteration $j$, $\left[\bm{\alpha}, \bm{\beta}\right]\left(j\right) $.
% we obtain the following recursive expression for the FIM based on Eq. (21) in \cite{tichavsky1998} (Note-using the results in Corollary 1, we can obtain the following recursion for each relay link independently)  :
%\begin{equation}
%\left[\J_t\right](j) = \left[\D_{t-1}^{22}\right](j) -\left[\D_{t-1}^{21}\right](j) \left(\left[\J_{t-1}\right](j)+\left[\D_{t-1}^{11}\right](j)\right)^{-1}\left[\D_{t-1}^{12}\right](j),
%\end{equation}
%where
\begin{equation}
%\begin{split}
\left[\D_{t-1}^{11}\right](j)  = -\exE \left[\bold \nabla_{\x_{t-1}}\left\{\bold \nabla_{\x_{t-1}} \log p\left(\x_t|\x_{t-1}\right)
\right)	^T					\right] = 
\left[
\begin{array}{ccc}
\frac{\left[\alpha\right](j)^2}{1- \left[\alpha\right](j)^2} & 0&0\\
0 & \frac{\left[\beta\right](j)^2}{1- \left[\beta\right](j)^2}&0\\
0 & 0&0
\end{array}%
\right] .
\end{equation}

\begin{equation}
%\begin{split}
\left[\D_{t-1}^{12}\right](j) = \left[\D_{t-1}^{21}\right](j)  = -\exE \left[\bold \nabla_{\x_{t-1}}\left\{\bold \nabla_{\x_{t-1}} \log p\left(\x_t|\x_{t-1}\right)
\right)	^T					\right] = 
\left[
\begin{array}{ccc}
\frac{\left[\alpha\right](j)}{1- \left[\alpha\right](j)^2} & 0&0\\
0 & \frac{\left[\beta\right](j)}{1- \left[\beta\right](j)^2}&0\\
0 & 0&0
\end{array}%
\right] .
%%%%%%%%%%%%%
\end{equation}
We obtain the following matrix decompositions of our system model, via Eqs. (34-36) of \cite{tichavsky1998}, 
\begin{equation}
\begin{split}
\left[\D_{t-1}^{22}\right](j)  &=
 -\exE \left[\bold \nabla_{\x_{t}}\left\{\bold \nabla_{\x_{t}} \log p\left(\x_t|\x_{t-1}\right)
\right)	^T\right]
-\exE \left[\bold \nabla_{\x_{t}}\left\{\bold \nabla_{\x_{t}} \log p\left(\y_t|\x_{t}\right)
\right)	^T\right]\\
 &= 
\left[
\begin{array}{ccc}
\frac{1}{1- \left[\alpha\right](j)^2} +\frac{\left|s\right|^2}{\sigma_v^2} & 0& \frac{s}{\sigma^2_v}\\
0 & \frac{1}{1- \left[\beta\right](j)^2}+\frac{\left|s\right|^2+\sigma^2_w}{\sigma_v^2}&0\\
\frac{s}{\sigma^2_v} & 0&\frac{1}{\sigma^2_w}+1
\end{array}%
\right] 
.
\end{split}
%%%%%%%%%%%%%
\end{equation}
In the simulation results Section we evaluate for each iteration of the PMCMC chain, $j$, the following: $\left[\text{Trace}\left(\frac{1}{T}\sum{t=1}^T F_{t}^{\left(j\right)}\left(\g_{t},\h_{t}, \w_{t}\right)\right)\right](j)$.
We produce distributional estimates of this quantity for the $J$ iterations of the Markov chain for different SNRs.\\
\textbf{Remark 5} - In our model framework we get analytic expressions for the BCRLB recursion. However, a key point about utilising this recursive evaluation for the FIM matrix is that in the majority cases one clearly can not evaluate the required expectations analytically over the joint distribution of the data and latent states. However, since we are constructing a particle filter proposal distribution for the AdPMCMC algorithm to target the filtering distribution $p\left(\x_t|\y_{1:t},[\bm{\theta}](j)\right)$ we can use this particle estimate to evaluate the expectations at each iteration $t$ for each data set. It is important to note that this recursion avoids ever calculating the expectations using the entire path space empirical estimate, only requiring marginal filter density estimates, which wont suffer from degeneracy as a path space empirical estimate would. 

%%%%%%%%%%%%%%%%%%%%%%%%%%%%%%%%%%%%%%%%%%%%%%%%%%%%%%%%%%%%%%%%%%%%%%%%%%%%%%%%
\section{Complexity Analysis} \label{complexity}
%%%%%%%%%%%%%%%%%%%%%%%%%%%%%%%%%%%%%%%%%%%%%%%%%%%
The complexity analysis of the class of algorithms presented in the previous sections can be studied from two perspectives;
the most technical of these involves theoretical study of the mixing rate of the MCMC algorithms under consideration,
the other focus would be on the computational complexity. 
Here we focus on a computational complexity comparison between each of the algorithms. The computational cost of each of these algorithms can be split into three parts:
the first cost involves constructing and sampling from the proposal;
the second significant computational cost comes from the evaluation of the acceptance probability for the proposed
new Markov chain state; and the third is related to the mixing rate of the overall MCMC algorithm as affected by the length
of the Markov chain required to obtain estimators of a desired accuracy.
We define the following building blocks for a single MCMC iteration and their associated complexity:
\begin{enumerate}
	\item  Sampling a random variable using exact sampling $\approx\mathbb{O}\left(1\right)$
	\item  Likelihood evaluation of $\prod_{n=1}^T \prod_{l=1}^L p \left(y^{(l)}_{n}|\alpha^{(l)}, \beta^{(l)},g_{n}^{(l)},h_{n}^{(l)} \right)$  $\approx TL\left(C_m+C_a\right)+\mathbb{O}\left(1\right)$
	\item  Prior evaluations of $\prod_{n=1}^T \prod_{l=1}^L 
														p \left(h_{n}^{(l)}|h_{n-1}^{(l)} \right)
														p \left(g_{n}^{(l)}|g_{n-1}^{(l)} \right)
														p \left(g_{1}^{(l)} \right)
														p \left(h_{1}^{(l)} \right)
														p \left(\alpha^{(l)} \right)
														p \left(\beta^{(l)} \right)$		  $\approx 6 TL \mathbb{O}\left(1\right)$
\end{enumerate}
Based on these building blocks we estimate the overall complexity of the proposed algorithms as follows.
\subsection{Construction and sampling of PMCMC proposal}
\subsubsection{Adaptive MCMC component}
(Step 3 of Algorithm \ref{Generic_PMCMC}) Complexity $\left(2L \times 2L+ 2L\right) \mathbb{O}\left(1\right)$

\subsubsection{Rao-Blackwellised SIR filter component}
(Step 4 of Algorithm \ref{Generic_PMCMC})
\begin{itemize}
	\item Kalman filter component : $TL \mathbb{O}\left(1\right)$.
	\item SIR filter component : $2NLT \mathbb{O}\left(1\right)$.
	\item Evaluation of marginal likelihood: $NT \mathbb{O}\left(1\right)$.
	\item Sampling SIR filter path space proposal:  $N \mathbb{O}\left(1\right)$.
\end{itemize}
\subsubsection{Evaluation of AdPMCMC acceptance probability}
(Step 7 of Algorithm \ref{Generic_PMCMC})
Complexity $ \left(NT+4L\right) \mathbb{O}\left(1\right)$.

Therefore, the total cost of a single AdPMCMC iteration can be approximated as\\  $ \left(2L^2+ TL+NT\left(2L+2\right)+N\right) \mathbb{O}\left(1\right)$.
%%%%%%%%%%%%%%%%%%%%%%%%%%%%%%%%%%%%%%%%%%%%%%%%

\subsection{Construction of MCMC proposal}
The total computational complexity of one iteration of the deterministic scan Gibb sampler, in Algorithm \ref{Standard_MHinGibbs}, involves $L\left(3T+2\right)$ parameters and requires updating and accepting each proposed move.
This produces $L\left(3T+2\right) \times \left(2TL+2\right) \mathbb{O}\left(1\right)= \left(6T^2L^2+10TL+4L\right)\mathbb{O}\left(1\right)$ operations.
%%%%%%%%%%%%%%%%%%%%%%%

%%%%%%%%%%%%%%%%%%%%%%%%%%%%%%%%%%%%%%%%%%%%%%%%%%%
\section{Results} \label{simulationresults}

The aim of this section is to demonstrate the performance of the AdPMCMC algorithm in performing channel tracking and estimation. To do this we separate the analysis into two sub-parts. We begin by analysing properties of the Adaptive MCMC and the Rao-Blackwellised particle filter proposal distribution in the context of the PMCMC algorithm. This involves analysis of Markov chain paths, the acceptance probability and the SIR filter performance.

These aspects will be studied under three different settings: the dimension of the posterior is increased by increasing the length of the frame $T$; the number of particles in the Rao-Blackwellised SIR filter, $N$, is increased; and the SNR is varied. In addition, we demonstrate the significant improvement in computation efficiency that our AdPMCMC algorithm has over MH-within-Gibbs.

The second part considers the wireless communications analysis. It aims to address the question of how well this proposed methodology solves the difficult problem of joint channel tracking and estimation by considering the estimated MSE of the channels and the distribution of the MMSE for the parameters of the non-linear state space models. 
%%%%%%%%%%%%%%%%%%%%%%%%%%%%%%%%%%%%
\subsection{Analysis of AdPMCMC algorithm performance versus $T$, $N$ and SNR}
In this section we evaluate the performance of joint channel tracking and parameter estimation under the AdPMCMC framework.
The network topology used in the simulations involved a single relay network, $K=L=1$. In all simulations we ran a Markov chain of length 50,000 iterations, discarding the first $15,000$ samples as burnin. We then systematically varied each of the three variables ($T,N$,SNR) and assessed the performance of the AdPMCMC algorithm. We took values of $N$ ranging from $10, 20, 50, 100, 200, 500, 1,000$ and 5,000. The length of frame considered involved $T$ ranging from 50, 100 and 200 symbols per frame, leading to posteriors to be sampled from in dimensions $152$, $302$ and $602$ respectively. Finally, we consider SNR levels from $0$ dB through to $25$dB which covers a wide range of possible operating environments.

The prior distributions for the parameters $\alpha$ and $\beta$ were identical and specified to have a Beta distribution, $Be(10,0.6)$. 
%The choice of this prior reflects the knowledge we have of the state space models in \ref{}. Firstly, the values of $\alpha$ and $\beta$ must be in the interval $[0,1]$ and this choice of prior satisfies this constraint. In addition, in the context of the relay tracking model, it is most sensible to consider values for $\alpha$ and $\beta$ closer to one. The reason for this is that under Jakes model, the values of $\alpha$ and $\beta$ are directly related to the physical velocity or speed at which the mobile transmitter at the source and the relay are moving. Values significantly less than one result in unrealistic speeds for the transmitter and relay. 
%We considered simulations in which the true value of $\alpha$ and $\beta$ ranged from fast fading with $\alpha = \beta = 0.8$ through to slow fading in which $\alpha = \beta = 0.95$.
\vspace{0.2cm}
\subsubsection{Analysis of the number of particles N versus acceptance rate of AdPMCMC algorithm} in this section we study the performance of the PMCMC algorithm average acceptance probability as a function of the number of particles N. This is meaningful as it allows us to recommend a setting for N. To achieve this we consider $\alpha = \beta = 0.95$ used to generate data corresponding to a flat fading channel model with an SNR of 15dB and frame length of $T=100$. 

\begin{figure} 
    \centering
        \epsfysize=5cm
        \epsfxsize=9cm
        \epsffile{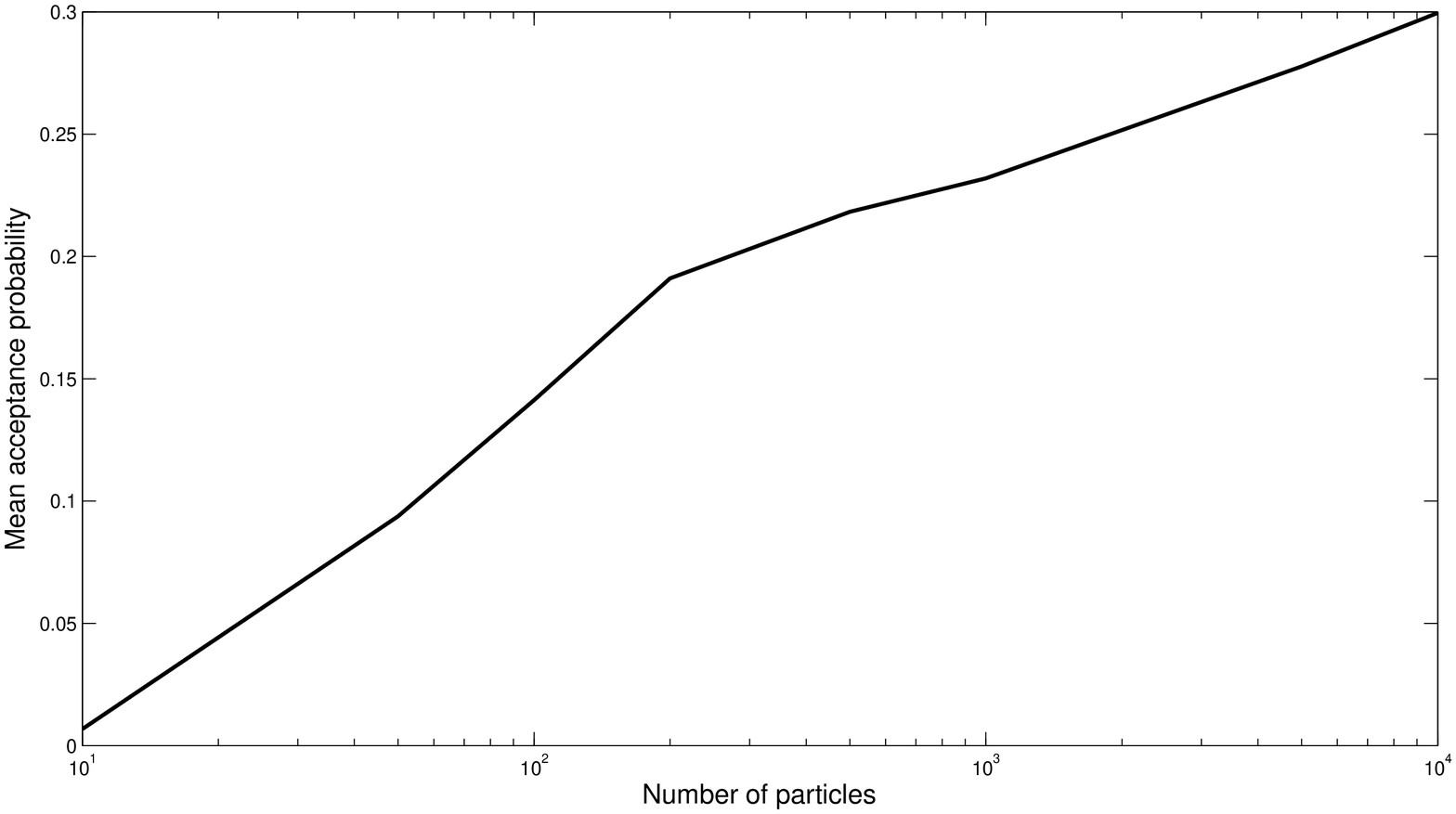}
        \caption{\textbf{AdPMCMC:} Acceptance probability for different number of particles}
    \label{fig:BER1}
\end{figure}

In Fig. \ref{fig:BER1} we present the average acceptance probability for the AdPMCMC algorithm for SNR=15dB, T=100, corresponding to a posterior distribution in $302$ dimensions. The key finding of this study is that we only require a very small number of particles to obtain accurate estimation and efficient performance in our PMCMC algorithm.
In particular we see that as expected, when the number of particles increases, the average acceptance probability of the joint proposal in the AdPMCMC Markov chain increases. Secondly, we note that even for a relatively small number of particles, $N=100$ we obtain average acceptance rates around 20\%. This is a very good indication that it is suitable to work with the Rao-Blackwellised SIR filter for our proposal. It is typical in the MCMC literature to tune a proposal mechanism in a standard MCMC algorithm to produce acceptance rates between [0.2, 0.5] and in some cases it is provably optimal to use $0.234$, see \cite{roberts2001optimal}. 

\subsubsection{Analysis of the estimated MMSE versus SNR}
\begin{figure} 
    \centering
        \epsfysize=6cm
        \epsfxsize=5cm
        \epsffile{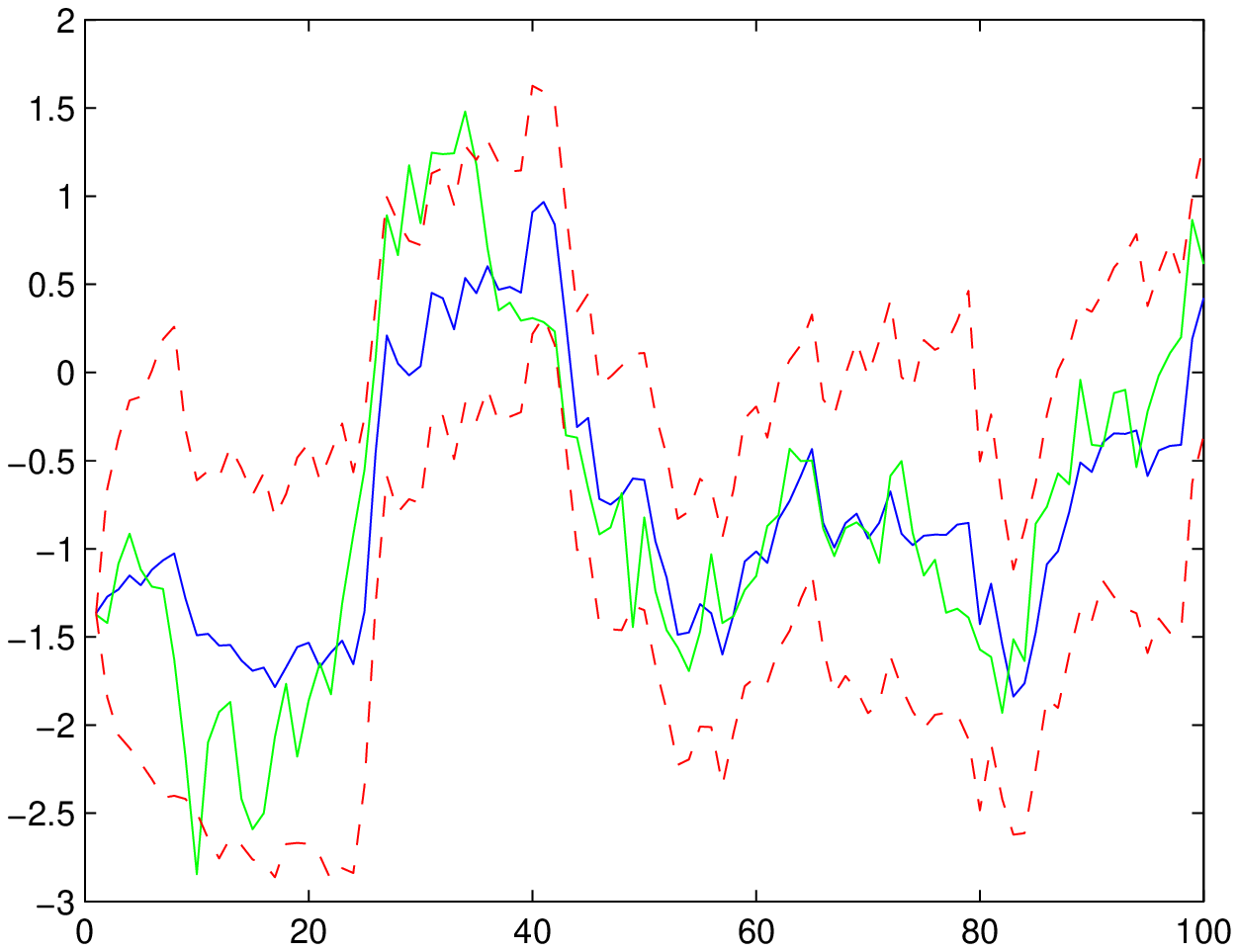}
        \epsfysize=6cm
        \epsfxsize=5cm
        \epsffile{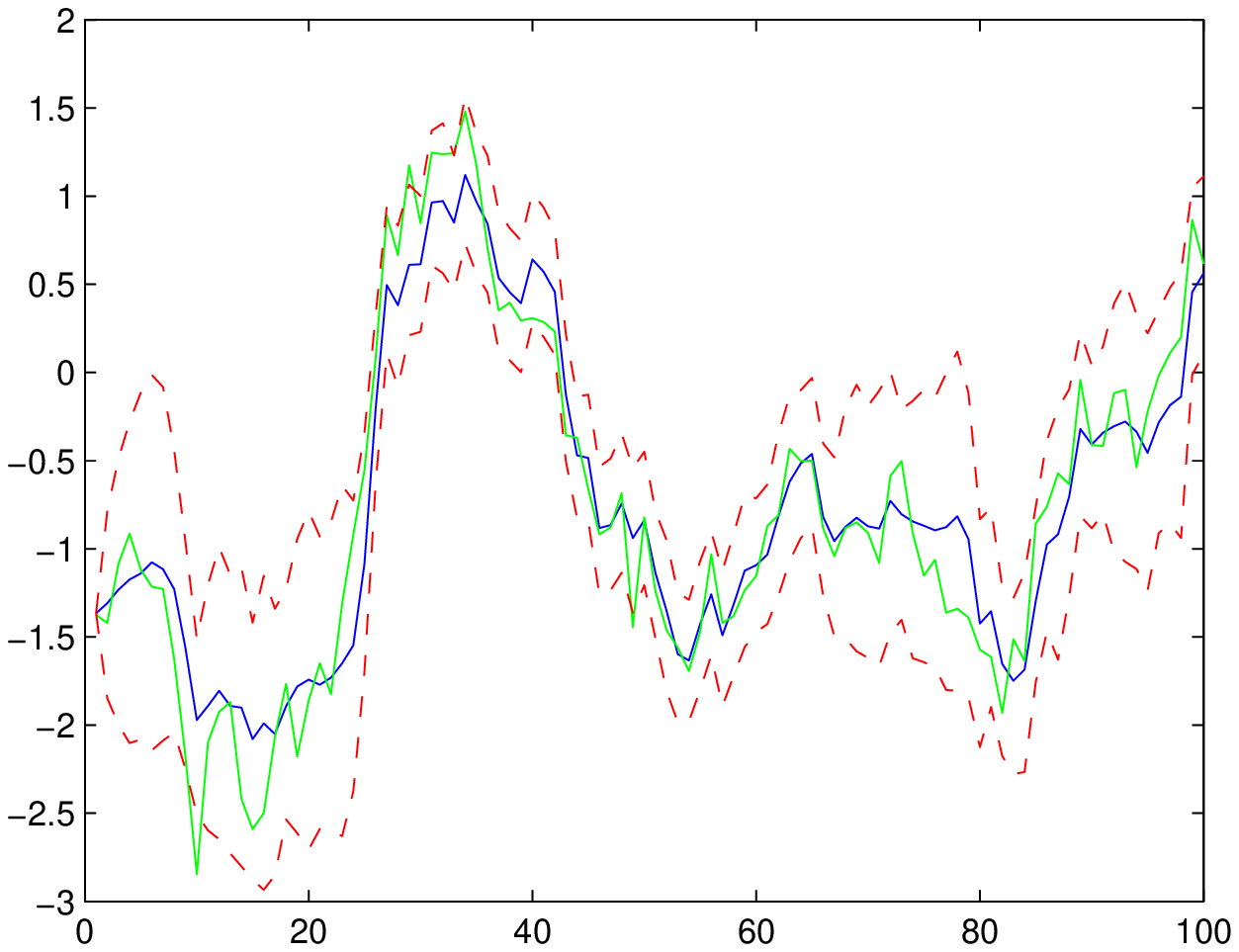}        
        \epsfysize=6cm
        \epsfxsize=5cm
        \epsffile{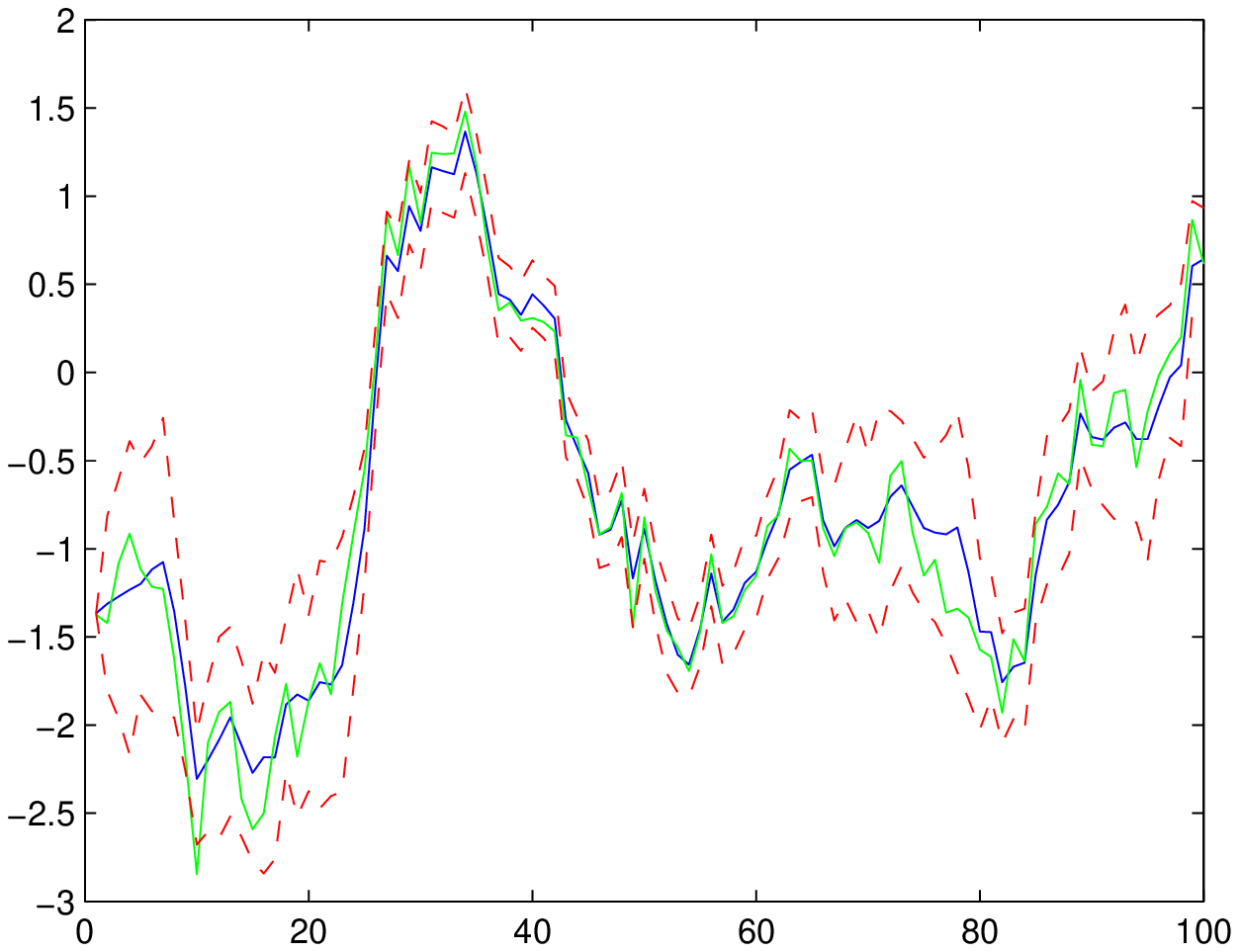}        
        \caption{\textbf{AdPMCMC:} MMSE for $\h$ with $95$\% posterior CI for low, medium and high SNRs, $T=100$, $N=100$. Note, true latent process is presented in green, the MMSE estimate in blue and posterior confidence intervals in dashed red line. }
    \label{fig:BER2}
\end{figure}

\begin{figure} 
    \centering
        \epsfysize=6cm
        \epsfxsize=5cm
        \epsffile{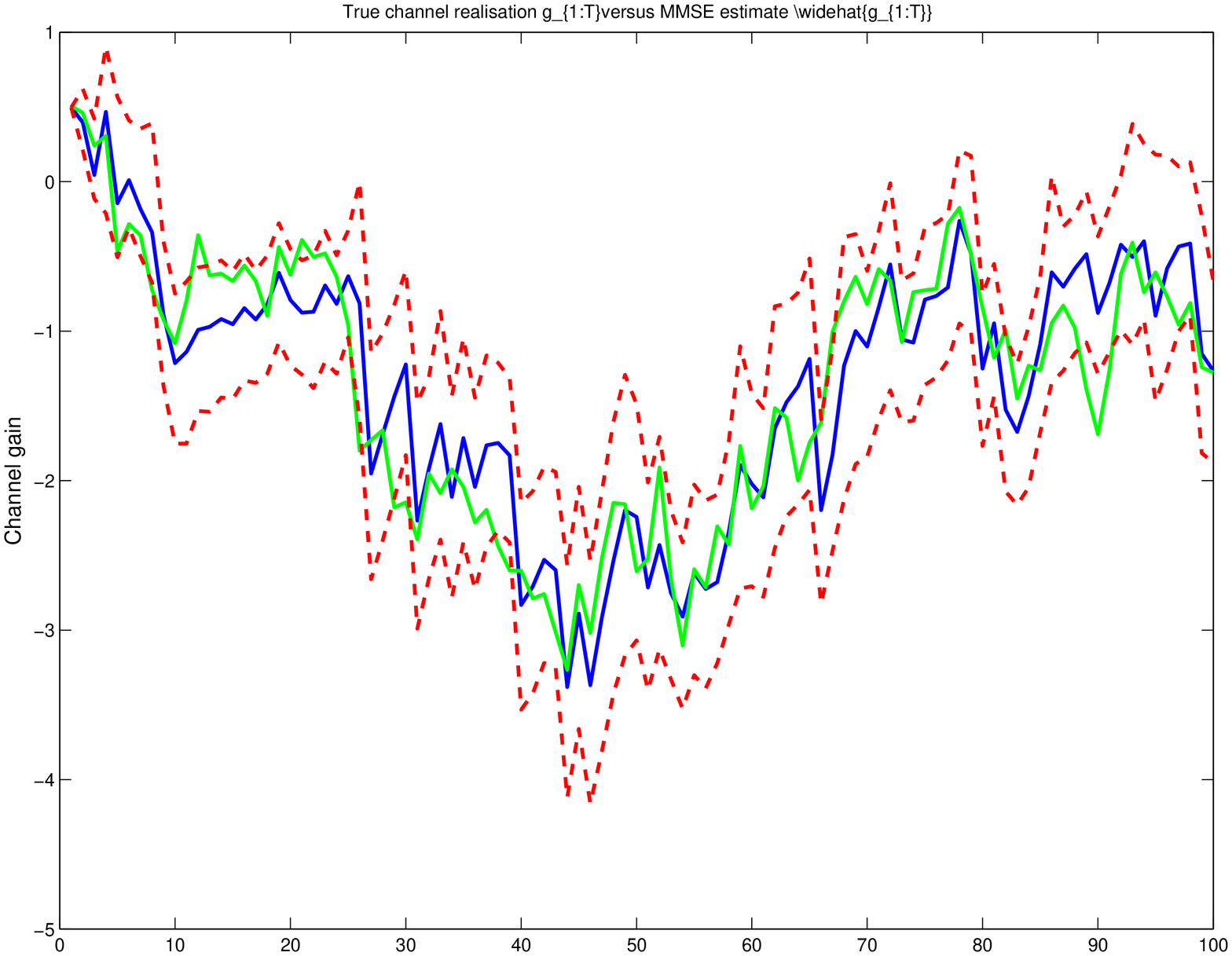}
        \epsfysize=6cm
        \epsfxsize=5cm
        \epsffile{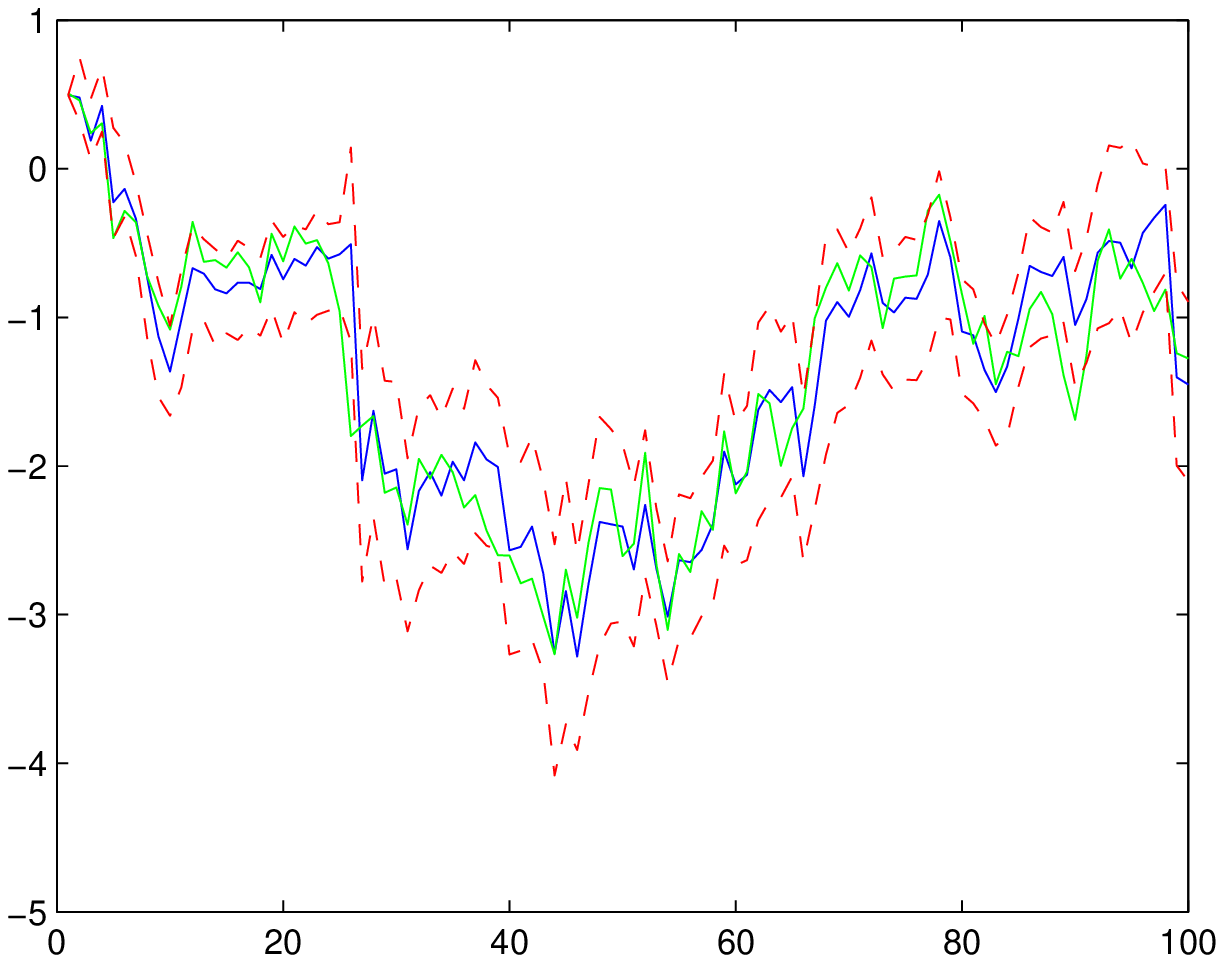}        
        \epsfysize=6cm
        \epsfxsize=5cm
        \epsffile{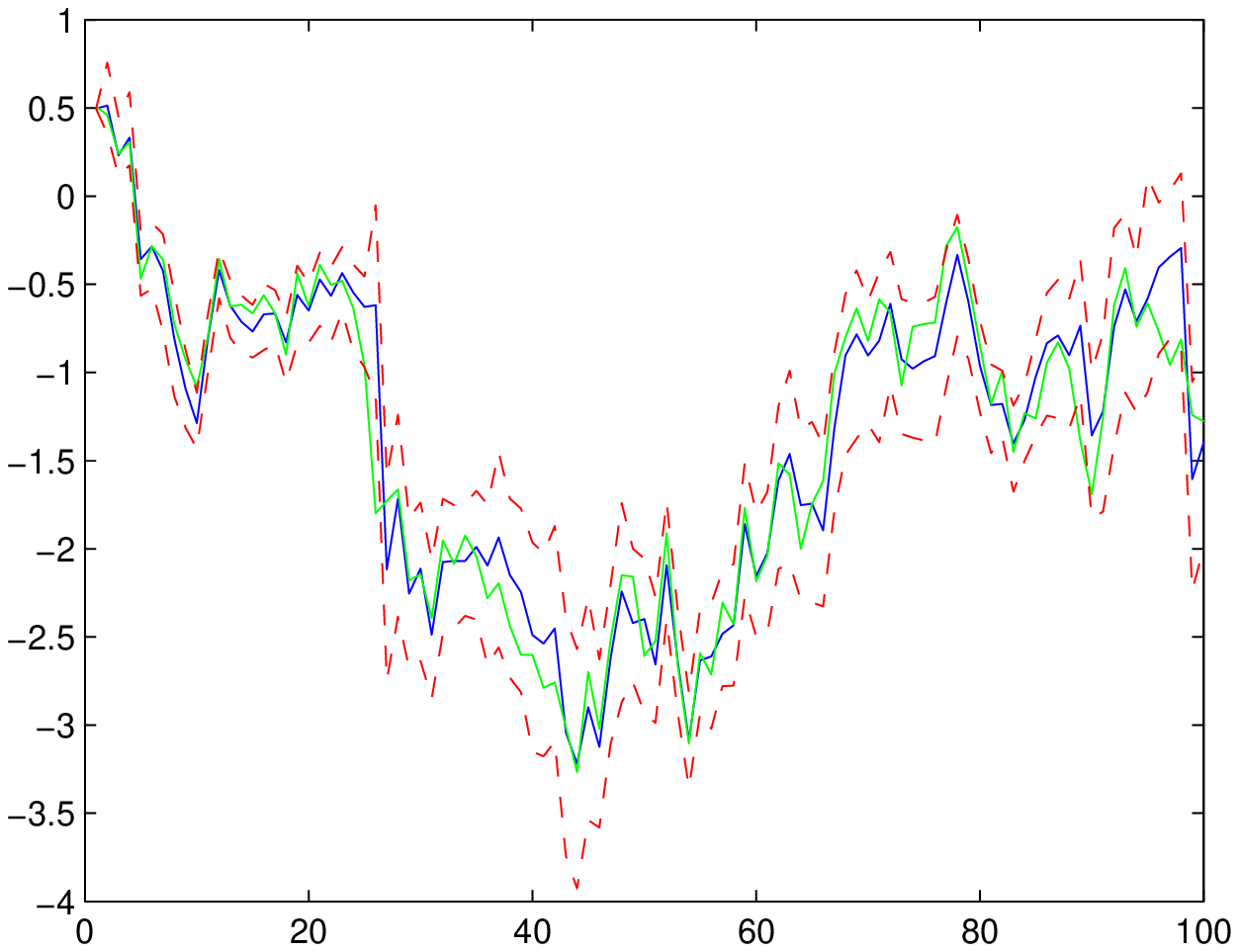}        
        \caption{\textbf{AdPMCMC:} MMSE for $\g$ with $95$\% posterior CI for low, medium and high SNRs, $T=100$, $N=100$.
        Note, true latent process is presented in green, the MMSE estimate in blue and posterior confidence intervals in dashed red line.
        }
    \label{fig:BER3}
\end{figure}

in this section we study the impact that the SNR will have on the estimation and channel tracking under the AdPMCMC algorithm with $N=100$, $T=100$, $\alpha = \beta = 0.95$ and the SNR ranging from low, medium and high corresponding to 0dB, 15dB and 25dB respectively. The sequence of subplots in Figs. \ref{fig:BER2} and \ref{fig:BER3} demonstrate the MMSE estimates of the channel estimations for a frame obtained from the AdPMCMC algorithm, versus the SNR. 

The first set of subplots demonstrates the improvement of the MMSE estimates of the channel $\h_{1:T}$ as the SNR increases. Additionally, we provide the $95\%$ posterior confidence intervals for the MMSE estimate on the path space, which demonstrates a reduction in uncertainty as the SNR increases. The second set of subplots demonstrate the same quantities for the estimation of channels $\g_{1:T}$ under this scenario. 

\begin{figure} 
    \centering
        \epsfysize=8cm
        \epsfxsize=10cm
        \epsffile{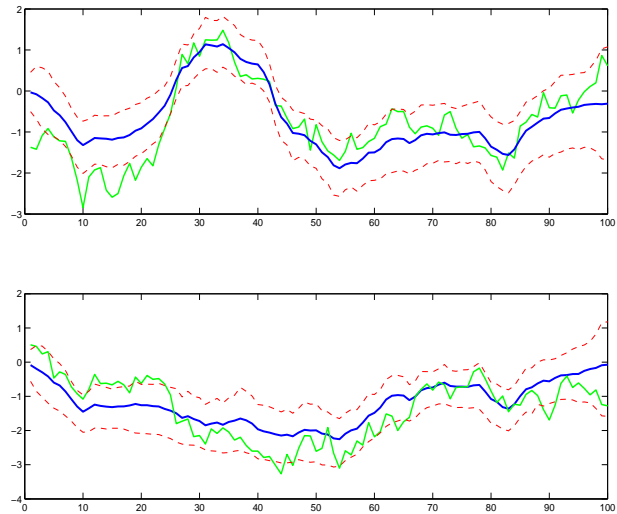}        
        \caption{\textbf{Gibbs:} MMSE for $\h$ (top panel) and $\g$ (lower panel) with $95$\% posterior CI for medium SNR, $T=100$.
        Note, true latent process is presented in green, the MMSE estimate in blue and posterior confidence intervals in dashed red line.}
    \label{fig:gibbs}
\end{figure}

Next we compare the MMSE results of the basic MH-within-Gibbs sampler of Algorithm \ref{Standard_MHinGibbs} to the MMSE results we obtained for the AdPMCMC.
In order to perform a fair comparison with respect to algorithmic complexity, detailed in Section \ref{complexity}, we set the length of the MH-within-Gibbs Markov chain to produce the same computational expense as the AdPMCMC sampler. This produced very poor results for the Gibbs sampler, and so, instead we present here an increase of computational cost of the Gibbs sampler by roughly $100$ times the length of the Markov chain.
In Fig. \ref{fig:gibbs} we present the MMSE estimates for the path space parameters $\h_{1:T}$ since $\g_{1:T}$. These results were obtained using the MH-within-Gibbs sampler with identical Markov chain initialisation to that used in the AdPMCMC.
In addition, we pre-tuned the MH-within-Gibbs sampler to have an acceptance rate of approximately $20 \%$.
Clearly, the poor mixing rate of the basic Gibbs sampler results in sub-optimal performance, requiring significantly longer chain to obtain the same accuracy as the AdPMCMC sampler. As a result, the proceeding analysis will continue with the AdPMCMC algorithm.

\begin{figure} 
    \centering
        \epsfysize=4.2cm
        \epsfxsize=4.2cm
        \epsffile{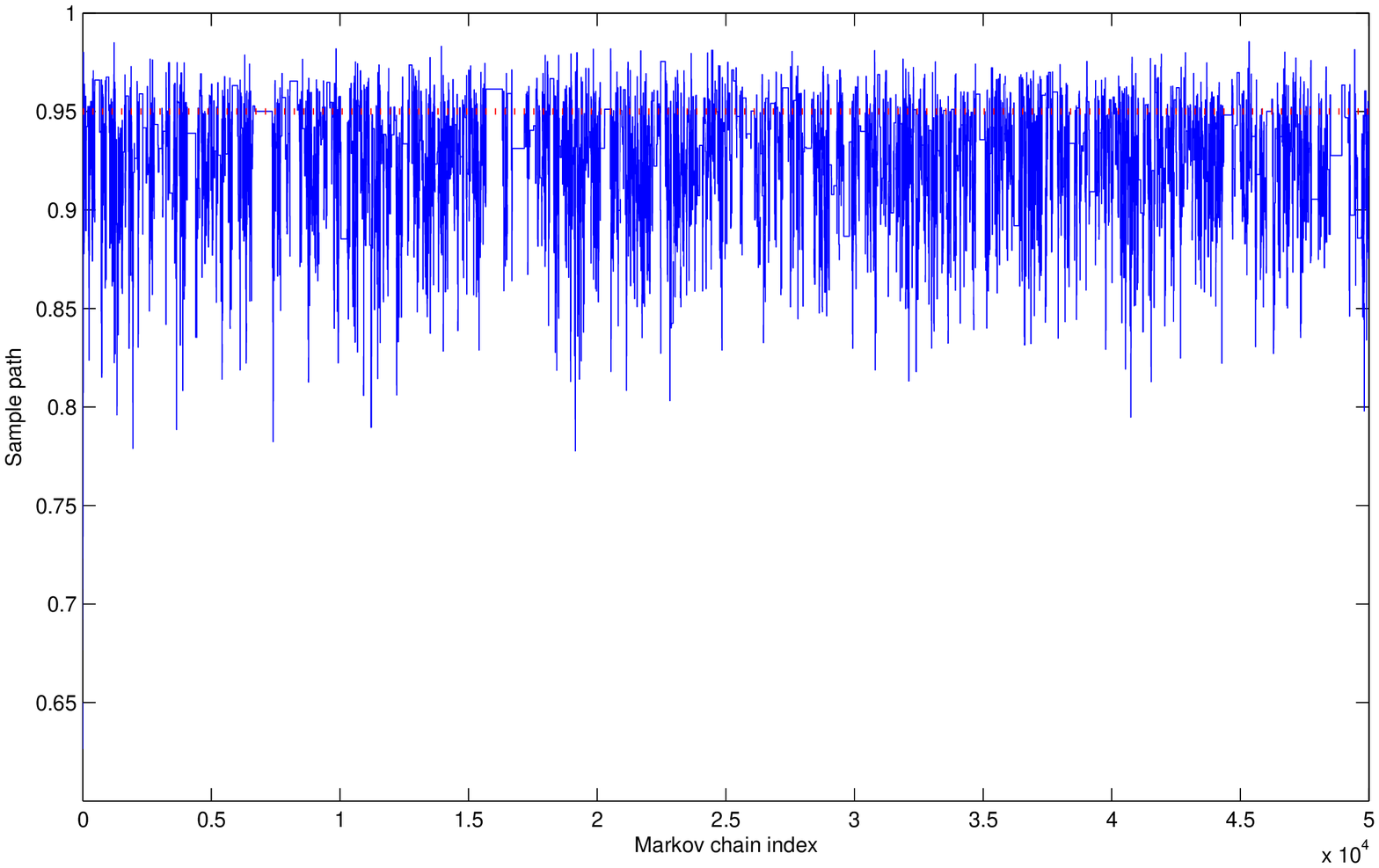}
        \epsfysize=4.2cm
        \epsfxsize=4.2cm
        \epsffile{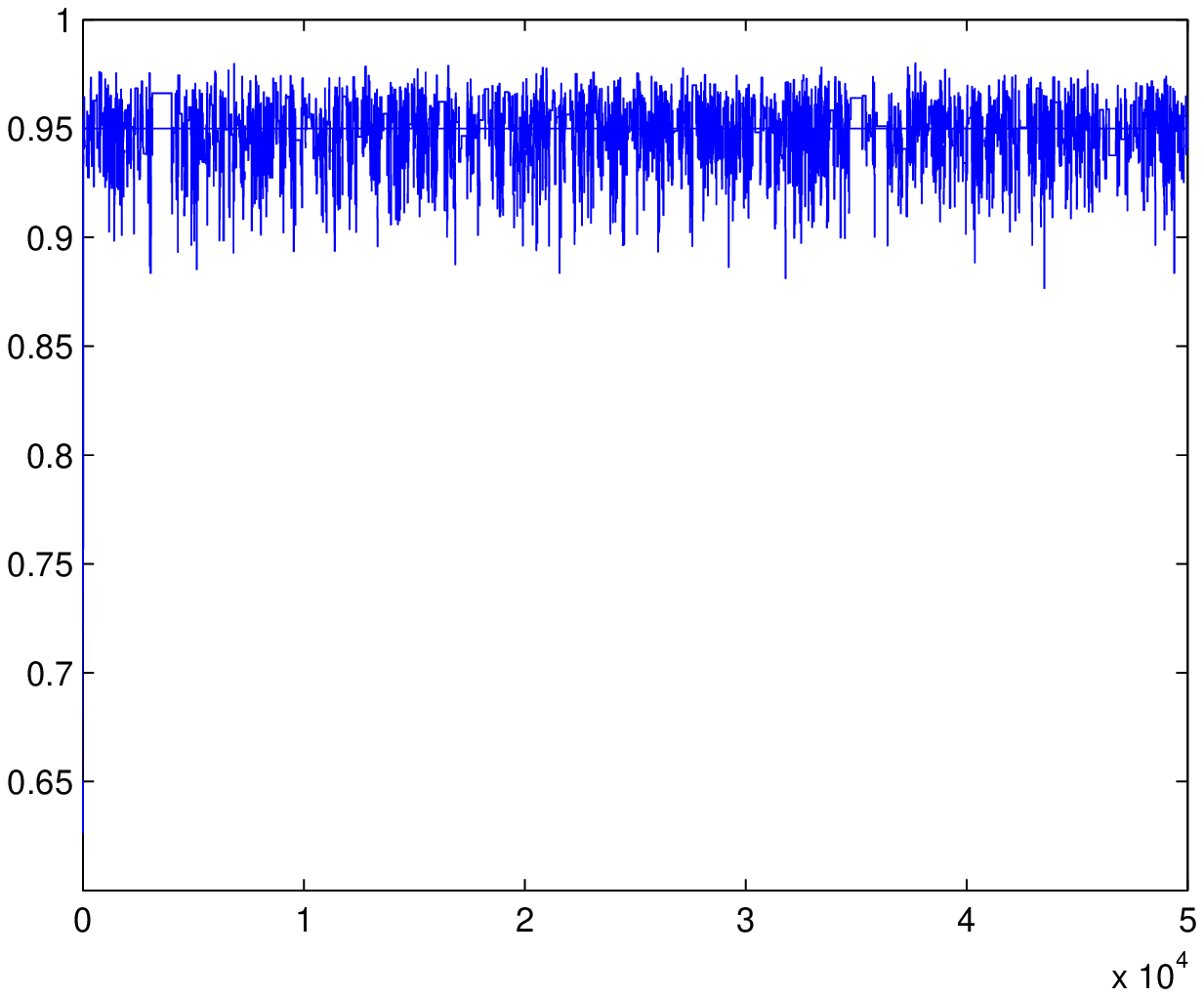}        
        \epsfysize=4.2cm
        \epsfxsize=4.2cm
        \epsffile{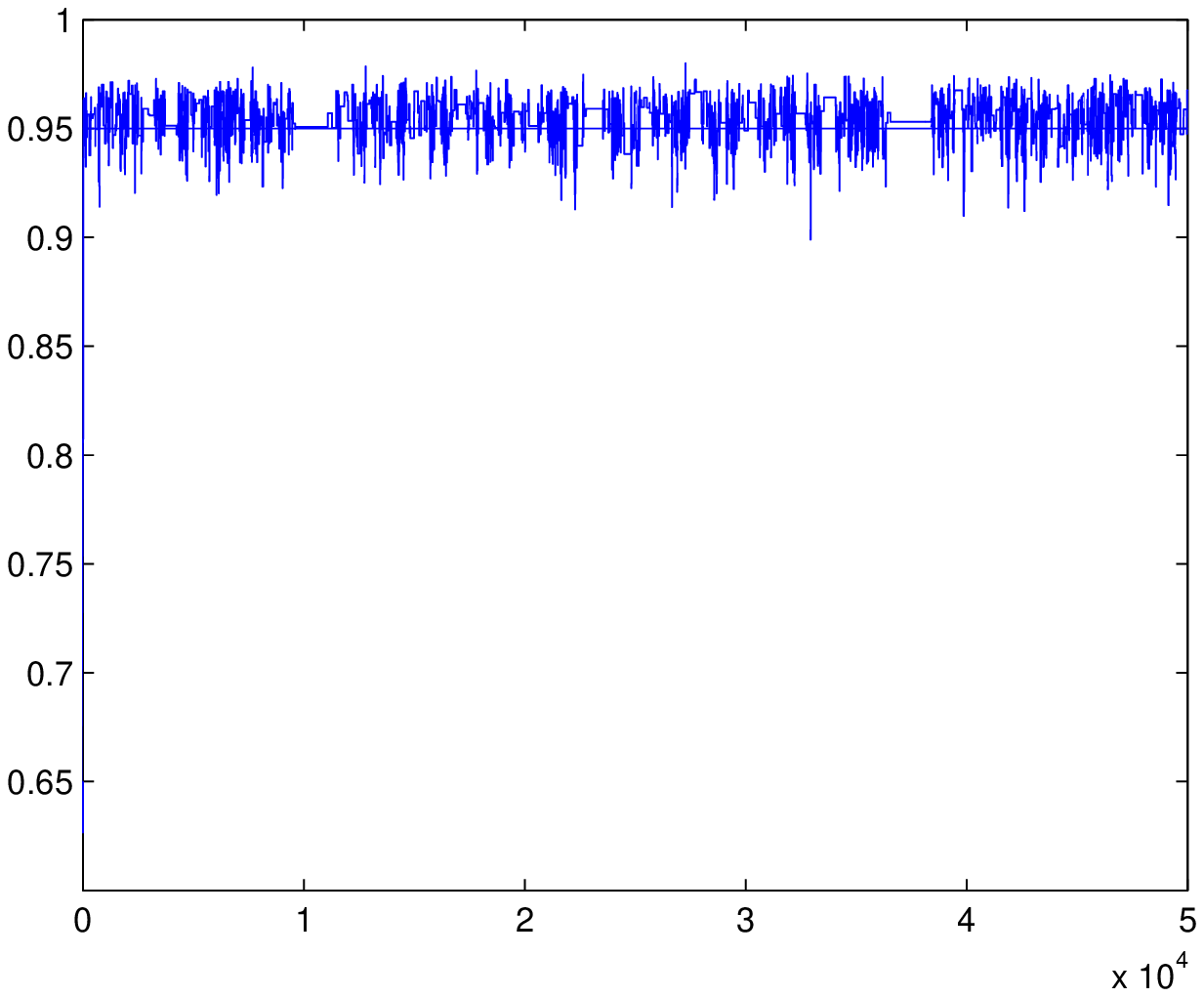}        
        \caption{\textbf{AdPMCMC:} sample path for $\alpha$  for low, medium and high SNRs, $T=100$, $N=100$}
    \label{fig:BER4}
\end{figure}

\begin{figure} 
    \centering
        \epsfysize=4.2cm
        \epsfxsize=4.2cm
        \epsffile{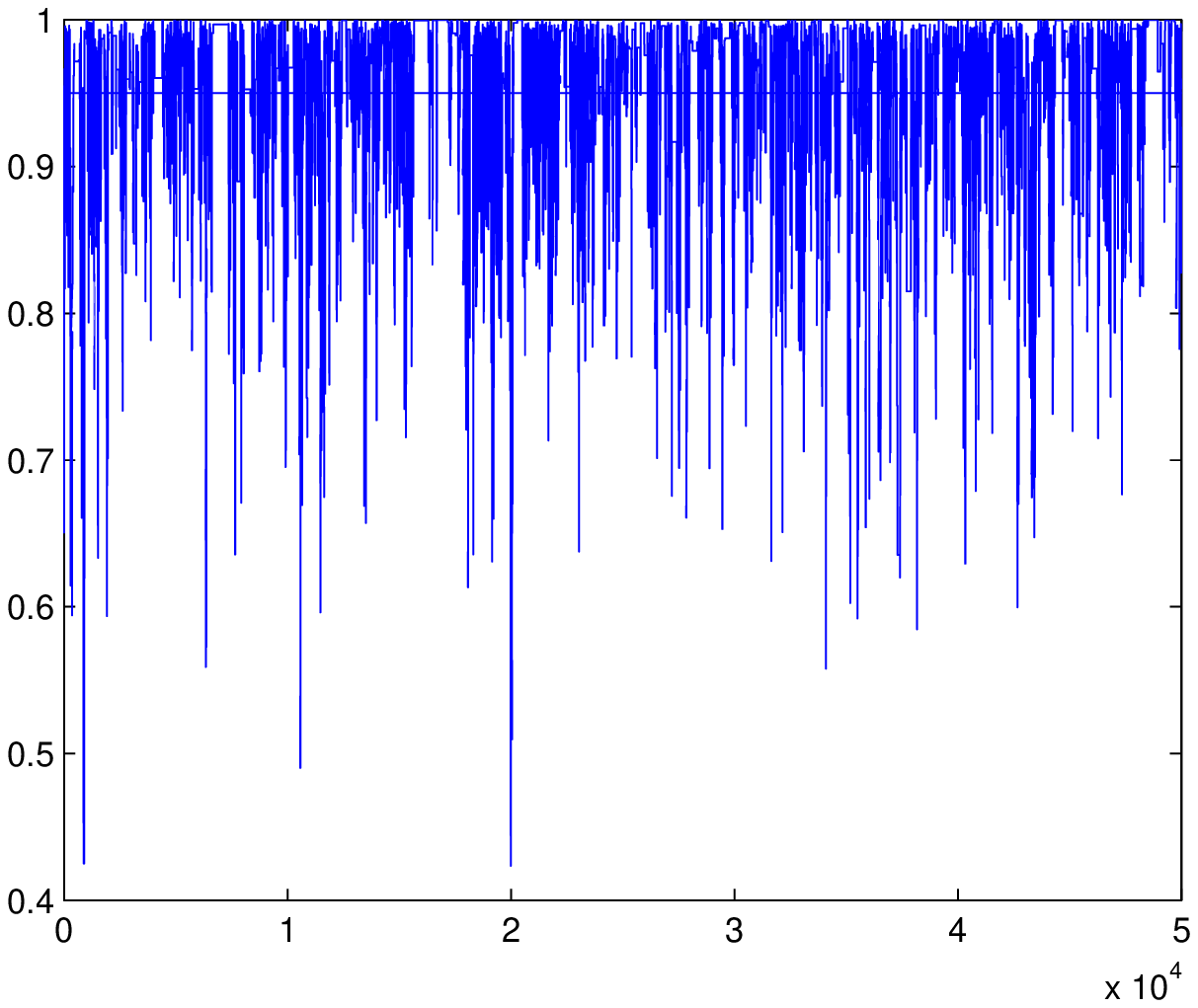}
        \epsfysize=4.2cm
        \epsfxsize=4.2cm
        \epsffile{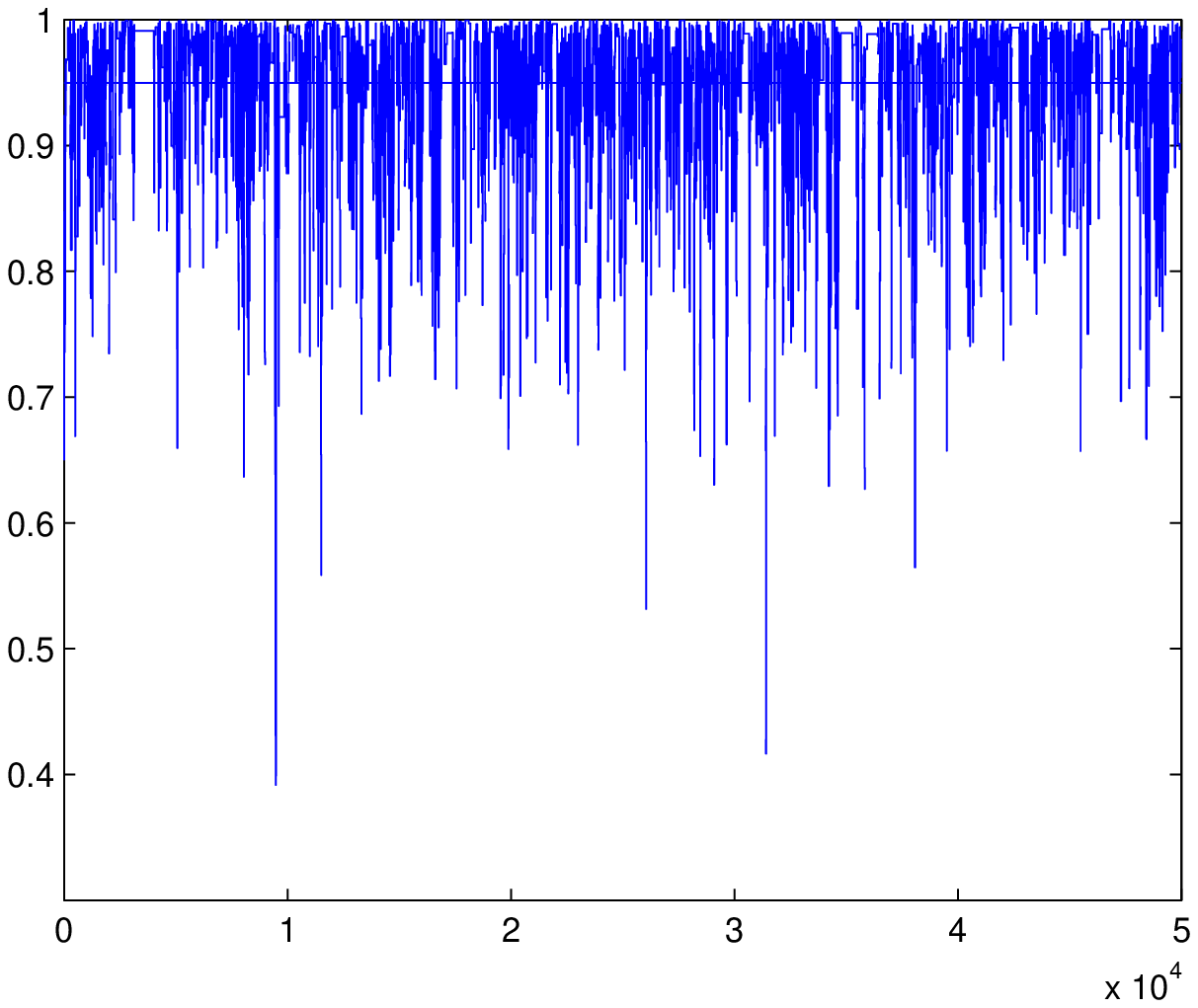}        
        \epsfysize=4.2cm
        \epsfxsize=4.2cm
        \epsffile{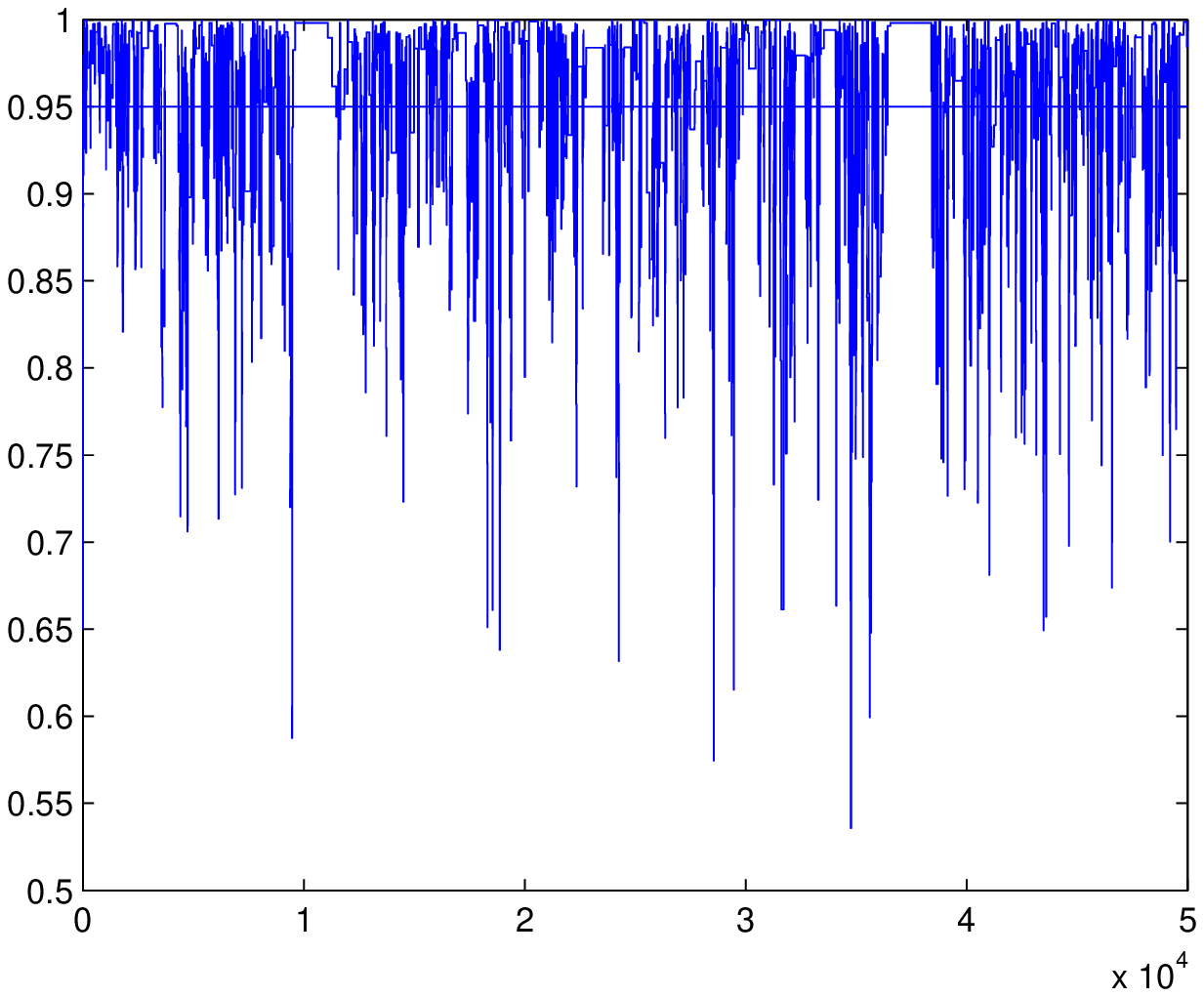}        
        \caption{\textbf{AdPMCMC:} sample path for $\beta$  for low, medium and high SNRs, $T=100$, $N=100$}
    \label{fig:BER5}
\end{figure}

Finally, we conclude this section with an analysis of the sample paths for the AdPMCMC Markov chain for the model parameters $\alpha$ and $\beta$. These are again presented as a function of the SNR level after having been obtained via the adaptive MH proposal mechanism designed to sample these components of the state vector at each iteration of the Markov chain. The results are presented in Figs. \ref{fig:BER4} and \ref{fig:BER5} for $\alpha$ and $\beta$ respectively. We see clearly that the precision of the posterior and therefore the estimation accuracy for $\alpha$ is strongly affected by the SNR level. This was not found to be the case for our model when considering $\beta$, clearly under the postulated model the posterior precision is not strongly affected by the SNR level. In addition we note that the MMSE estimates of the parameters for $\alpha$ and $\beta$ was highly accurate at around 0.95 for all SNR. This is strong support of our proposed AdPMCMC algorithm for performing the joint estimation and channel tracking efficiently.

\subsection{Analysis of estimated MSE for the AdPMCMC algorithm channel estimates versus SNR}

\begin{figure} 
\centering
\subfigure[] 
{   
    \includegraphics[width=0.42\textwidth]{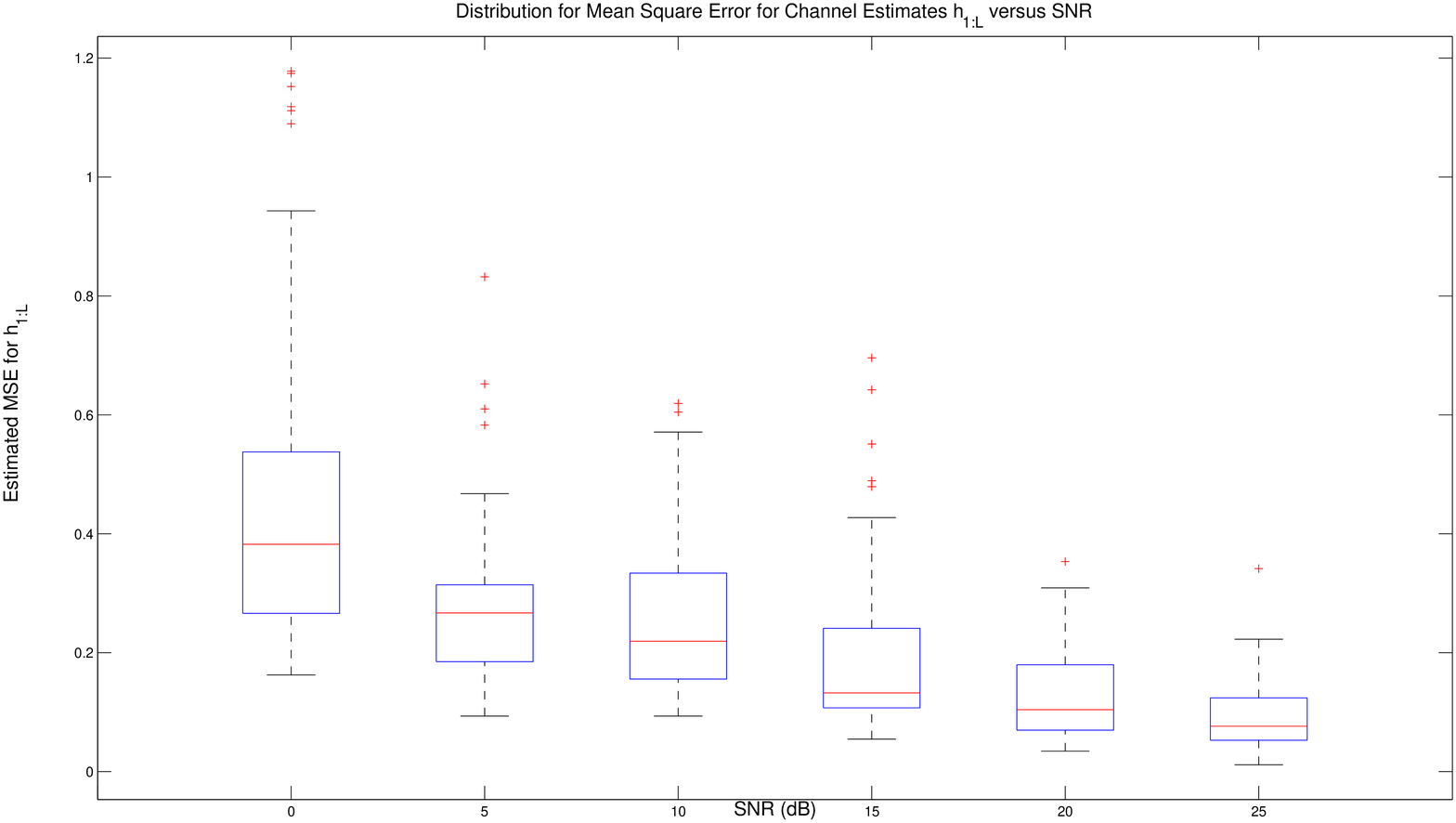}
}
\subfigure[] 
{    
    \includegraphics[width=0.42\textwidth]{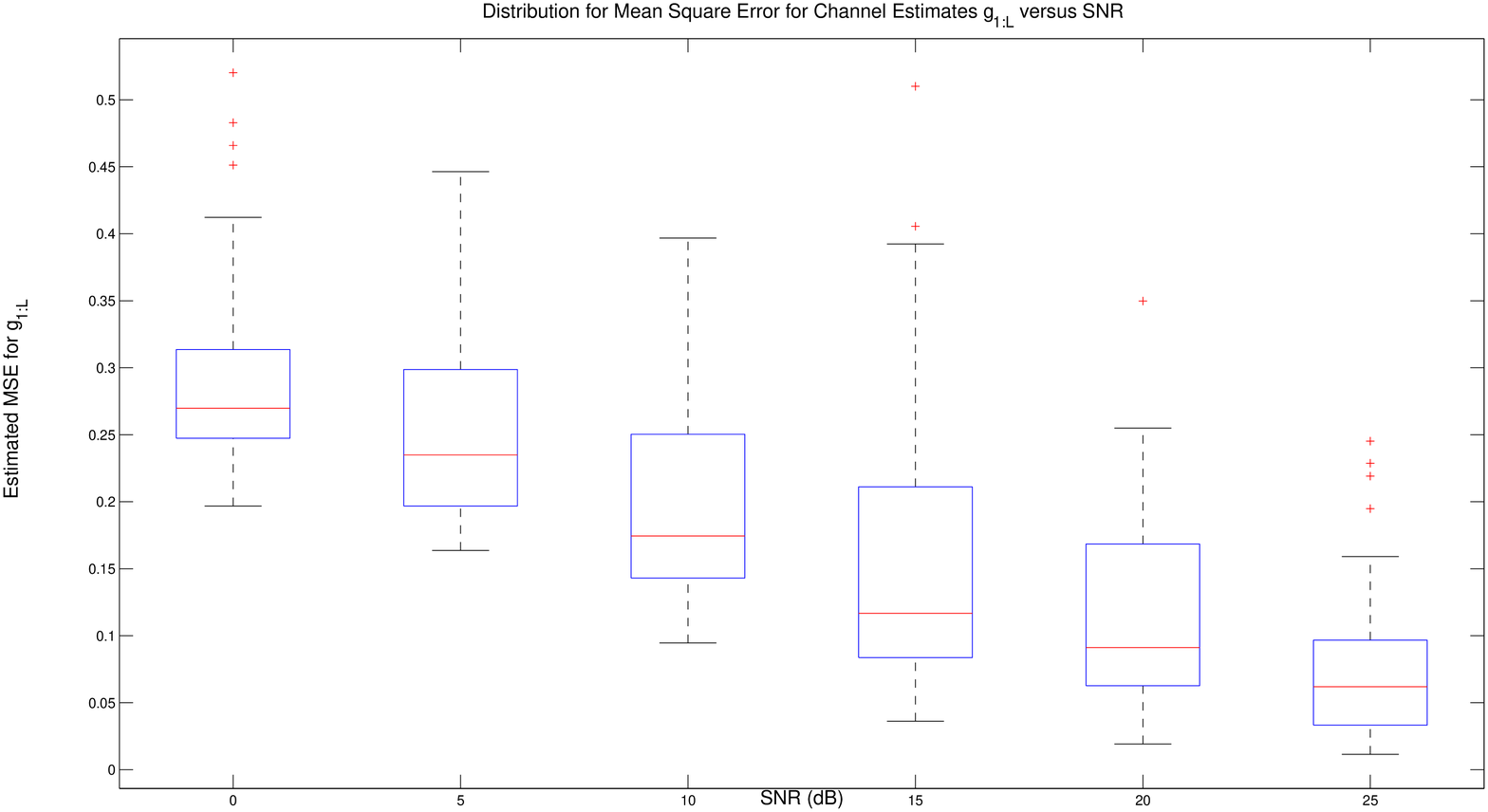}
}
 \caption{Distribution of estimated MSE of $\h$ (left panel) and $\h$ (right panel) vs. SNR over $500$ frames, T=$100$, N=$100$. The solid line represents the average MSE vs SNR.}
\label{fig:BER6} 
\end{figure}

In this section we consider an experiment in which the SNR is varied from 0dB to 25dB in increments of 5dB. For each SNR level, $500$ frames of data transmission are generated in which each frame is of length $T=100$. The number of relays present is set to $L=1$ and the true values of $\alpha$ and $\beta$ used to generate the observations for each frame were equal at 0.95. For each data transmission the AdPMCMC algorithm is performed for 25,000 iterations, with the first 10,000 samples discarded as burnin. Then using the remaining Markov chain paths we obtained an estimate of the MMSE for the channels $\h_{1:T},\g_{1:T}$ for each of the transmitted frames. We then obtain the total MSE for each frame and plot the box-whisker plot of the distribution of the MSE as a function of SNR for the transmitted frames, under our AdPMCMC estimation approach.
The results of this analysis are presented in Fig. \ref{fig:BER6}. Clearly, there is a decrease in the Total MSE as the SNR increases. In addition we note that as expected, since the estimate of the channels $\g_{1:T}$ is performed using the Rao-Blackwellising conditionally optimal Kalman Filter, this is reflected in the level of the total MSE. The estimates for $\g_{1:T}$ are clearly more accurate than those for the particle filter sampled estimates $\h_{1:T}$.

\begin{figure} 
\centering
\subfigure[] 
{   
    \includegraphics[width=0.42\textwidth]{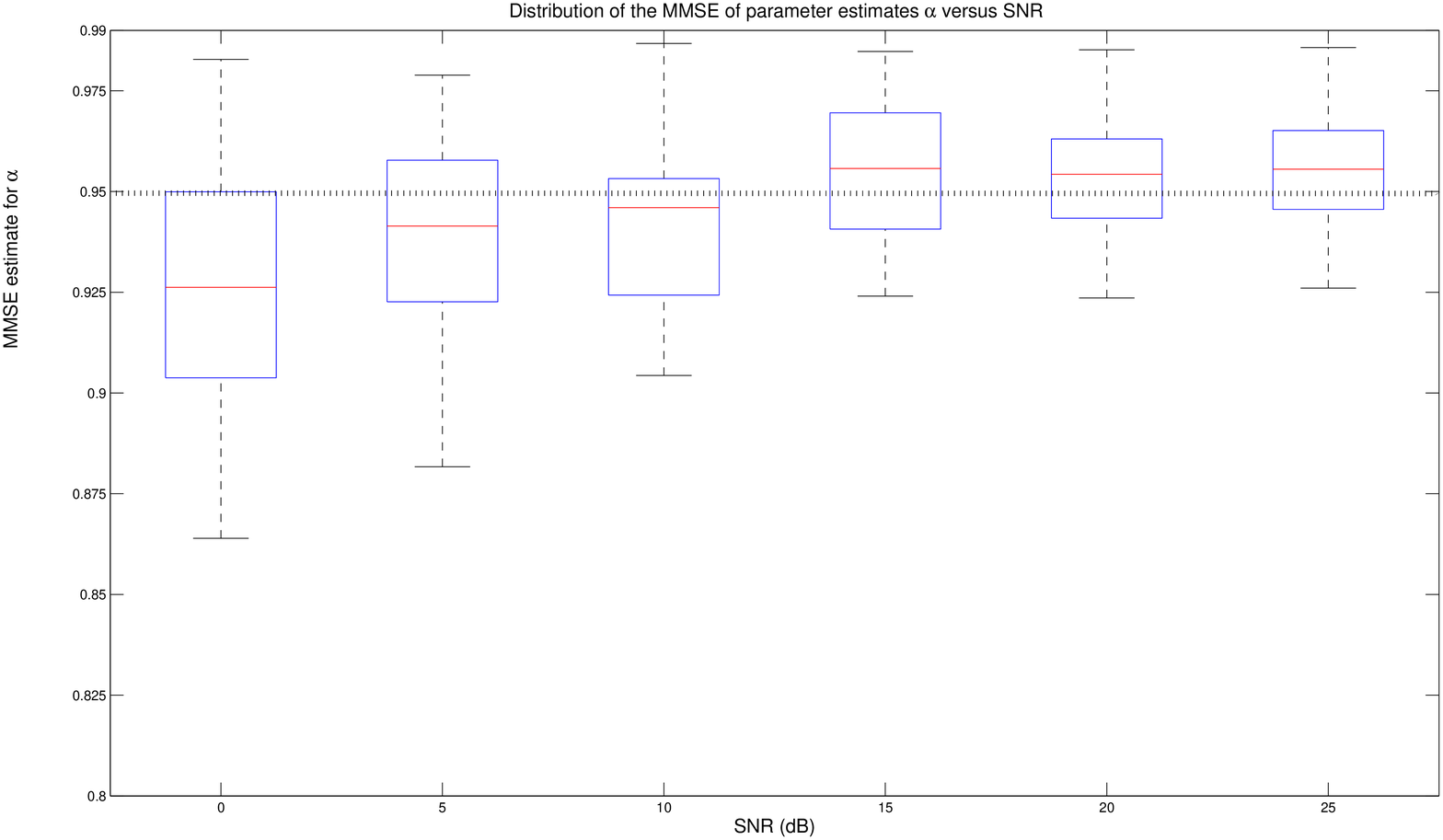}
}
\subfigure[] 
{    
    \includegraphics[width=0.42\textwidth]{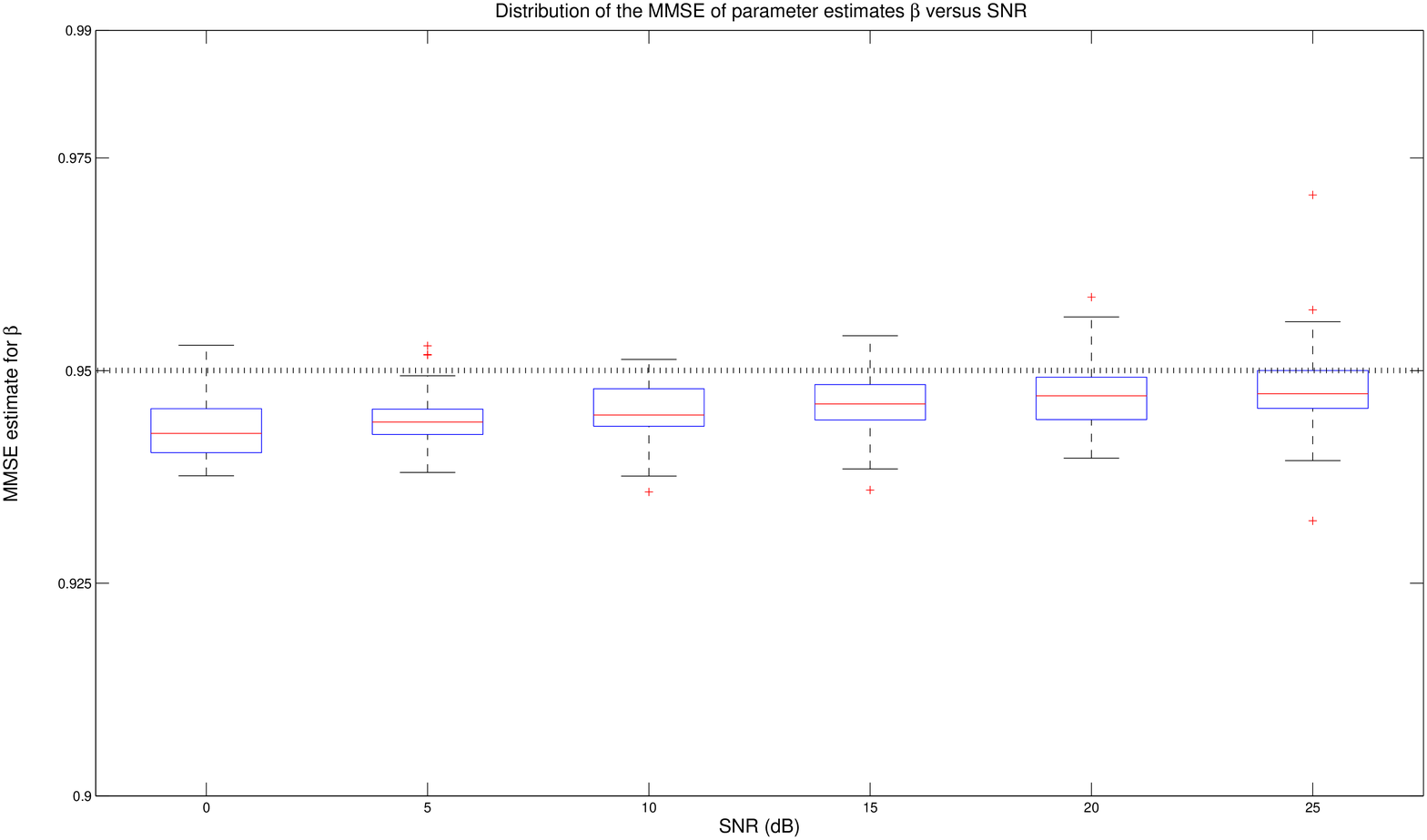}
}
 \caption{boxplot MMSE for $\alpha$ (left panel) and $\beta$ (left panel) vs. SNR over 40 frames, T=100, N=100.}
\label{fig:BER8} 
\end{figure}

In addition we present the distribution of the estimated MMSE for the parameters $\bm{\alpha}$ and $\bm{\beta}$ over the $500$ frames as a function of the SNR. Clearly as the SNR increases, the estimates converge to the true parameter values and in addition the precision in the MMSE estimates over each independent frame, reduces. This clearly appears in Fig. \ref{fig:BER8}.

\begin{figure} 
\centering
\subfigure[BCRLB for $\g_{1:T}$] 
{
    \label{fig:BCRLB_g}
    \includegraphics[width=0.42\textwidth]{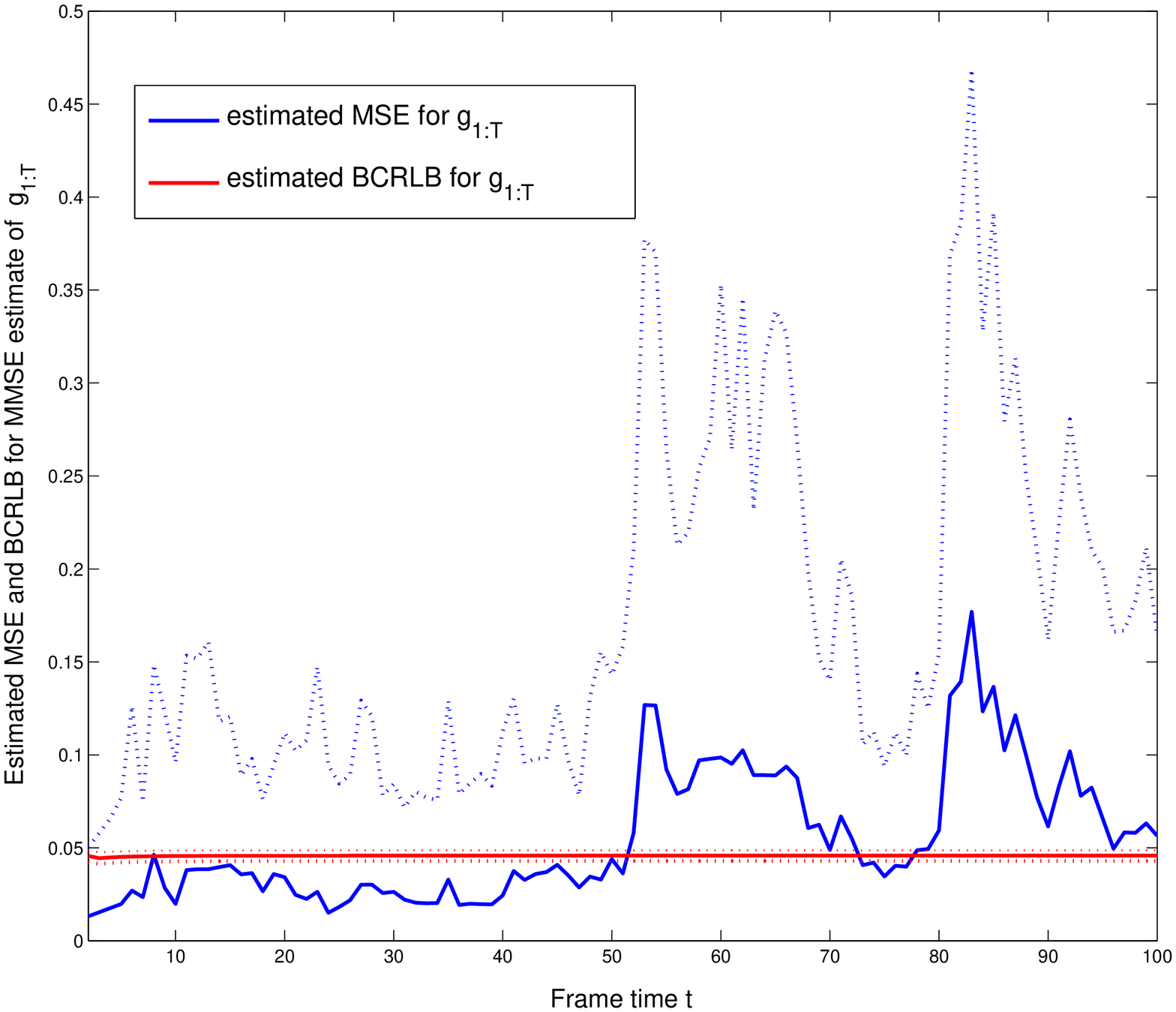}
}
\subfigure[BCRLB for $\h_{1:T}$] 
{
    \label{fig:BCRLB_h}
    \includegraphics[width=0.42\textwidth]{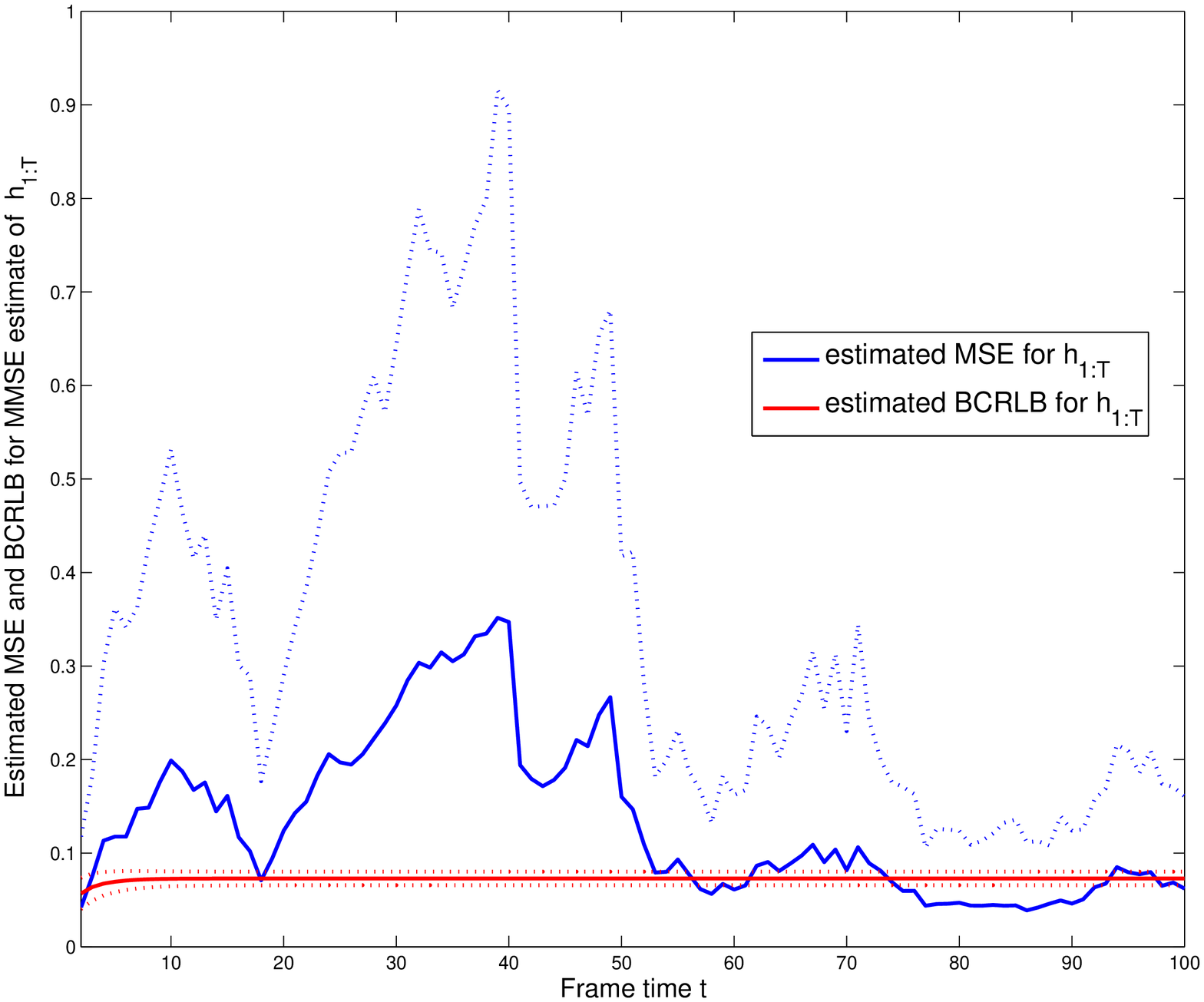}
}
 \caption{BCRLB estimated after marginalizing out uncertainty from model parameters $\bm{\alpha},\bm{\beta}$ and averaged over $500$ data realizations. Each PMCMC Markov chain was simulated for J = $25,000$ iterations.}
\label{fig:BCRLB} 
\end{figure}

In Fig. \ref{fig:BCRLB} we present the path space estimate for the BCRLB after marginalizing out the model parameters according to eq. (\ref{MC_CRLB}). This is presented for an SNR=15dB after simulation of the PMCMC algorithm for $500$ independently generated data sets, for PMCMC chains each of length J=25,000 iterations. We also present the estimated average MSE for the MMSE estimation of $h_{1:T}$ and $g_{1:T}$ and three standard deviations for this average MSE estimate at each step in the path space for these simulations. For each data set the MSE is estimated by splitting the Markov chain into subblocks of length 1,000 samples on which the MMSE is estimated. This estimate is then used to form an estimate of the MSE for the chain corresponding to the particular data set, finally these results are averaged over the data sets. The results demonstrate that the estimate of BCRLB is always close to the mean and easily within 3 standard deviations, indicating again that our PMCMC methodology and associated channel estimates perform close to optimally. 

%%%%%%%%%%%%%%%%%%%%%%%%%%%%%%%%%%%%%%%%%%%%%%%%%%%%%%%%%%%%%%%%%%%%%%%%%%%%%%%%
\section{Conclusions} \label{conclusions}
%%%%%%%%%%%%%%%%%%%%%%%%%%%%%%%%%%%%%%%%%%%%%%%%%%%%%%%%%%%%%%%%%%%%%%%%%%%%%%%%
We introduced a new approach for channel tracking and online parameter estimation in cooperative wireless relay networks. We developed a novel algorithm to efficiently solve the problem of joint channel tracking and parameters estimation within a mobile wireless relay network. This is based on a novel version of the PMCMC sampler. Simulation results demonstrate the effectiveness of the proposed algorithm in very high dimensional state-space. Though not the topic of this paper, we note that the proposed estimation methods can be trivially extended to also cover frequency selective channels.

%%%%%%%%%%%%%%%%%%%%%%%%
%%%%%%%%%%%%%%%%%%%%%%%
\bibliographystyle{IEEEtran}
\bibliography{../../../../references}
\vspace{0.5cm}

%%%%%%%%%%%%%%%%%%

%%%%%%%%%%%%%%%%%%

\begin{algorithm}
 \caption{Deterministic Scan Gibb Sampler.}
 \label{Standard_MHinGibbs}
     \begin{algorithmic}[1]
\STATE Set initial state $[\bm{\alpha}, \bm{\beta}, \g_{1:T}, \h_{1:T}, \w_{1:T}](0)$ deterministically or by sampling the priors;
%%%%%%%%%%%%%%%%%
\LINE {Repeat\;{$j$= $1$ to $J$}}

\STATE Sample $[\bm{\alpha}, \bm{\beta}](j+1) \sim  p \left(\bm{\alpha}, \bm{\beta}| [\g_{1:T},\h_{1:T},\w_{1:T}](j), y^{(l:L)}_{1:T} \right)$ 
%via a Metropolis-Hastings proposal, to obtain updated block of parameters which are accepted according to a Metropolis within Gibbs acceptance probability.
\hspace{0.5cm} 
\LINE {Repeat\;{$k=1$ to $K$}}
\STATE Sample 
$[\G_{(k-1)\tau+1:k\tau}](j+1) \sim p \left(\g_{(k-1)\tau+1:k\tau}|[\bm{\alpha}, \bm{\beta}, \g_{1:(k-1)\tau}](j+1),[\g_{k\tau+1:T\tau},\h_{1:T},\w_{1:T}](j), y^{(l:L)}_{1:T} \right)$ 

\hspace{0.5cm} 
\LINE {Repeat\;{$k=1$ to $K$}}
\STATE Sample $[\h_{(k-1)\tau+1:k\tau}](j+1) \sim p \left(\h_{(k-1)\tau+1:k\tau}|[\bm{\alpha}, \bm{\beta}, \h_{1:(k-1)\tau},\g_{1:T}](j+1),[\h_{k\tau+1:T\tau},\w_{1:T}](j), y^{(l:L)}_{1:T} \right)$
% from a Metropolis-Hastings proposal, to obtain updated block of parameters which are accepted according to a Metropolis within Gibbs acceptance probability which produces a block of samples with target full conditional posterior distribution: 
\hspace{0.5cm} 
\LINE{Repeat\;{$k=1$ to $K$}}
\STATE Sample $[\w_{(k-1)\tau+1:k\tau}](j+1) \sim p \left(\w_{(k-1)\tau+1:k\tau}| [\bm{\alpha}, \bm{\beta}, \w_{1:(k-1)\tau},\h_{1:T}, \g_{1:T}](j+1), [\w_{k\tau+1:T\tau}](j), y^{(l:L)}_{1:T} \right)$ 
\end{algorithmic}
\end{algorithm}

%%%%%%%%%%%%%%%%%%5
\begin{algorithm}[t]
\begin{algorithmic}[1]
\LINE \hspace{-1cm} \textit{Initialise Markov chain state:}
\STATE Initialise $\left[\bm{\alpha}, \bm{\beta}, \h_{1:T},\g_{1:T},\w_{1:T}\right](1)$ by sampling each value from the corresponding priors, see section \ref{ModelAssumptions}.
\LINE \hspace{-1cm} \textit{Begin PMCMC iterations:}
\FOR{$j=1,\hdots, J$}
\LINE \hspace{-1cm}\textit{Sample static parameters for $\bm{\alpha}, \bm{\beta}$ from adaptive mixture proposal, see Algorithm \ref{adaptice_MCMC}}
\STATE{Sample $\left[\bm{\alpha}, \bm{\beta}\right]^*(j+1)
 \sim q\left(\left[\bm{\alpha}, \bm{\beta}\right](j);\left[\bm{\alpha}, \bm{\beta}\right](j+1)\right) $.}
\LINE \hspace{-1cm} \textit{Construct and sample path space realisations for $\h_{1:T},\g_{1:T},\w_{1:T}$ from Rao-Blackwellised SIR filter proposal, see Algorithm \ref{SIR_filter} }
\STATE{Run an SMC algorithm with $N$ particles to obtain:

\begin{align}
\label{marginal_likelihood}
\begin{split}
\widehat{p}\left(\h_{1:T},\g_{1:T},\w_{1:T}|\y_{1:T},\left[\bm{\alpha}, \bm{\beta}\right]^*(j+1)\right)
&=\sum_{i=1}^{N} \left[\Xi_{1:T}\right](j+1,i)
\delta_{\left[\h_{1:T},\g_{1:T},\w_{1:T}\right](j+1,i)}\left(\h_{1:T},\g_{1:T},\w_{1:T}\right)\\
\widehat{p}\left(\y_{1:T}|\left[\bm{\alpha}, \bm{\beta}\right]^*(j+1)\right) &= \prod_{t=1}^T
\left(\frac{1}{N}\sum_{i=1}^{N} \left[\xi_{t}\right](j+1,i)
\right)
\end{split}
\end{align}
}%
\STATE Sample proposed path space
 $\left[\h_{1:T},\g_{1:T},\w_{1:T}\right]^*(j+1) \sim \widehat{p}\left(\h_{1:T},\g_{1:T},\w_{1:T}|\y_{1:T},\left[\bm{\alpha}, \bm{\beta}\right]^*(j+1)\right)$
 \LINE \hspace{-1.5cm} \textit{Combine proposals and calculate PMCMCM acceptance probability}
\STATE Combine both sampled proposals to obtain $\left[\bm{\alpha}, \bm{\beta},\h_{1:T},\g_{1:T},\w_{1:T}\right]^*(j+1)$
\STATE{Accept the proposed new Markov chain state comprised of
$\left[\bm{\alpha}, \bm{\beta},\h_{1:T},\g_{1:T},\w_{1:T}\right]^*(j+1)$ with acceptance probability
given by
\begin{align}
\begin{split}
&A \left(\left[\bm{\alpha}, \bm{\beta},\h_{1:T},\g_{1:T},\w_{1:T}\right](j);
\left[\bm{\alpha}, \bm{\beta},\h_{1:T},\g_{1:T},\w_{1:T}\right]^*(j+1)\right)
\\
 &= \min \left( 1, \frac{\widehat{p}\left(\y_{1:T}| \left[\bm{\alpha}, \bm{\beta}\right]^*(j+1) \right)
 p\left(\left[\bm{\alpha}, \bm{\beta}\right]^*(j+1)\right)
 q\left(\left[\bm{\alpha}, \bm{\beta}\right]^*(j+1); 
 \left[\bm{\alpha}, \bm{\beta}\right](j)
 \right)}
 {\widehat{p}\left(\y_{1:T}|\left[\bm{\alpha}, \bm{\beta}\right](j) \right)
 p\left(\left[\bm{\alpha}, \bm{\beta}\right](j)\right)
 q\left(\left[\bm{\alpha}, \bm{\beta}\right](j);\left[\bm{\alpha}, \bm{\beta}\right]^*(j+1)\right)
 }
 \right),
\end{split} 
\end{align}
where $\widehat{p}\left(\y_{1:T}|\left[\bm{\alpha}, \bm{\beta}\right](j)\right)$ is obtained
from the previous iteration of the PMCMC algorithm.}
%%%%%%%%%%%%%%%%%%5
\STATE Sample a realisation $u_1$ of random variable $U_1 \sim U[0,1]$
\IF {$u_1 < A \left(\left[\bm{\alpha}, \bm{\beta},\h_{1:T},\g_{1:T},\w_{1:T}\right](j)
\left[\bm{\alpha}, \bm{\beta},\h_{1:T},\g_{1:T},\w_{1:T}\right]^*(j+1)\right)$}
\STATE Accept proposed state $\left[\bm{\alpha}, \bm{\beta},\h_{1:T},\g_{1:T},\w_{1:T}\right](j+1)=\left[\bm{\alpha}, \bm{\beta},\h_{1:T},\g_{1:T},\w_{1:T}\right]^*(j+1)$ for iteration $j+1$
\ELSE
\STATE $\left[\bm{\alpha}, \bm{\beta},\h_{1:T},\g_{1:T},\w_{1:T}\right](j+1) = \left[\bm{\alpha}, \bm{\beta},\h_{1:T},\g_{1:T},\w_{1:T}\right](j)$
\ENDIF
%%%%%%%%%%%%%%%%%5
\ENDFOR
\caption{Generic PMCMC algorithm}
\label{Generic_PMCMC}
\end{algorithmic}
\end{algorithm}

\begin{algorithm}
\begin{algorithmic}[1]

\STATE Sample a realisation $u_1$ of random variable $U_1 \sim U[0,1]$
\LINE \hspace{-1cm}\textit{ Sample from the adaptive mixture proposal (\ref{AMCMC_random_walk}) }
\IF {$u_1 \geq w_1$}
\LINE \hspace{-1cm}\textit{ Sample $\left[\bm{\alpha}, \bm{\beta}\right](j+1)$ from the adaptive component of the mixture proposal of (\ref{AMCMC_random_walk})}
\STATE Estimate $\Sigma_j$, the empirical covariance of $\alpha, \beta$, using samples $\{[\bm{\alpha},\bm{ \beta}](i)\}_{i=1:j}$.\;
\STATE Sample proposal $\left[\bm{\alpha}, \bm{\beta}\right](j+1) \sim N
\left(\bm{\alpha}, \bm{\beta};\left[\bm{\alpha}, \bm{\beta}\right](j) ,\frac{\left(2.38\right)^2}{d}\Sigma_j\right)$;
\ELSE
\LINE \hspace{-1cm}\textit{ Sample $\left[\bm{\alpha}, \bm{\beta}\right](j+1)$ from the non-adaptive component of the mixture proposal of (\ref{AMCMC_random_walk})}
\STATE Sample proposal $\left[\bm{\alpha}, \bm{\beta}\right](j+1) \sim N\left(\bm{\alpha}, \bm{\beta};\left[\bm{\alpha}, \bm{\beta}\right](j),
\frac{\left(0.1\right)^2}{d}I_{d,d}    \right)$
\ENDIF

\caption{Adaptive MCMC for static parameters $\alpha, \beta$}
\label{adaptice_MCMC}
\end{algorithmic}
\end{algorithm}
%%%%%%%%%%%%%%%%%%%%%%%

\begin{algorithm}
\begin{algorithmic}[1]
\LINE \hspace{-1cm}\textit{Initialisation of Rao-Blackwellised SIR filter at iteration $(j+1)$ of the Markov chain}
\STATE{SIR particle filter for $\h_{1:T}, \w_{1:T}$: initialise $N$ particles $\left\{\left[\h_1, \w_1\right](j+1,i)\right\}_{i=1:N}$ via sampling from the priors in Section \ref{ModelSection}.}
\STATE{Rao-Blackwellised Kalman filter for $\g_{1:T}$: initialise the mean and covariance of $\g_1$ for each particle denoted by $\left\{\left[\mu_1,\Sigma_1\right](j+1,i)\right\}_{i=1:N}$}

\FOR{$t=2,\ldots, T$}
\LINE \hspace{-1cm}\textit{Perform mutation of the $N$ particles at frame time $t-1$ to obtain new particles at $t$ via state evolution.}
\STATE{Sample the $i$-th particle $\left[\h_t\right](j+1,i)$ from particle filter proposal \\
$\sim \mathcal{CN}\left( \left[\bm{\alpha}\right](j+1)   \left[\h_{t-1}\right](j+1,i),
 \sqrt{1- \left( \left[\bm{\alpha}\right](j+1)\right)^2}\right)$, 
according to state equation (\ref{Jakes})  .}

\STATE{Sample the $i$-th particle $\left[\w_t\right](j+1,i)$ from the prior, see Section \ref{ModelSection}.}

\LINE \hspace{-1cm}\textit{Perform Kalman filter evolution for sufficient statistics of $\g_n$.}
\STATE  For each particle, to obtain $\left\{\left[\mu_n,\Sigma_n\right](j+1,i)\right\}_{i=1:N}$, apply recursions in algorithm \ref{Kalman_filter}.

\LINE \hspace{-1cm}\textit{Incremental SIR importance sampling weight correction.}

\STATE{Evaluate the unnormalised importance sampling weights, $\left[\widetilde{\Xi}_t\right](j+1,i)$, for the $N$ particles, with the $i$-th weight given by
\begin{align}
\begin{split}
\left[\widetilde{\Xi}_t\right](j+1,i) &\propto  \left[\Xi_{t-1}\right](j+1,i) \left[\xi_{t-1}\right](j+1,i)\\
&\propto  \left[\Xi_{t-1}\right](j+1,i) p (\y_{t}|\left[\h_t,\g_t,\w_t,\bm{\alpha}, \bm{\beta}\right](j+1,i)),
\end{split}
\end{align}
}
\STATE{Normalise the importance sampling weights $\left[\xi_{t}\right](j+1,i)=
\frac{\left[\xi_{t-1}\right](j+1,i)}{\sum_{i=1}^{N}\left[\xi_{t-1}\right](j+1,i)}$}

\LINE \hspace{-1cm}\textit{Evaluate the importance estimate and resample adaptively.}

\STATE Calculate the Effective Sample size, $ESS =\frac{1}{\sum_{i=1}^{N}\left[\xi_{t-1}\right](j+1,i)^2} $
\STATE{If the Effective Sample size is less than 80\% then resample the particles at time $t$ using stratified resampling
based on the empirical distribution constructed from the importance weights to obtain new particles with equal weight.}
\ENDFOR
\STATE Evaluate marginal likelihood $\widehat{p}\left(\y_{1:T}|\left[\bm{\alpha}, \bm{\beta}\right](j+1)\right) = \prod_{t=1}^T
\left(\frac{1}{N}\sum_{i=1}^{N} \left[\xi_{t}\right](j+1,i)\right) $
\caption{Construction of proposal 
$\widehat{p}\left(\h_{1:T},\g_{1:T},\w_{1:T}|\y_{1:T},\left[\bm{\alpha}, \bm{\beta}\right](j+1)\right)$, given $\left[\bm{\alpha}, \bm{\beta}\right](j+1)$}

\label{SIR_filter}
\end{algorithmic}
\end{algorithm}
%%%%%%%%%%%%%%%%%%%%%%%
%%%%%%%%%%%%%%%%%%%%%%%
%%%%%%%%%%%%%%%%%%%%%%
\begin{algorithm}
\begin{algorithmic}[1]

 \caption{Kalman filter recursion for $\left[\g_n\right](j,i)$ particle $i$} 
     
\STATE sample $ \omega \sim \mathcal{CN}\left(0,1\right)$     
\STATE Predict;
%%%%%%%%%%%%%%%%%
\begin{align*}
&\left[\bm{\mu}_{n|n-1}\right](j,i) = \left[\bm{\mu}_{n-1|n-1}\right](j,i) + \sqrt{1- \left(\left[\bm{\beta}\right](j)\right)^2} \;\omega\\
&\left[\Sigma_{n|n-1}\right](j,i) =
\left[\bm{\mu}_{n-1|n-1}\right](j,i) \left[\Sigma_{n-1|n-1}\right](j,i) \left[\bm{\mu}_{n-1|n-1}\right](j,i)^{\top} +
 \sqrt{1- \left(\left[\bm{\beta}\right](j)\right)^2} 
\end{align*}
\STATE Update
\begin{align*}
& \widetilde{\y} = \y_n - f\left(s_n  \left[\h_n\right](j,i) ,n\right)\left[\bm{\mu}_{n|n-1}\right](j,i)\\
& \S =   f\left(s_n  \left[\h_n\right](j,i) ,n\right)\left[\bm{\mu}_{n|n-1}\right](j,i)
				 \left[\Sigma_{n|n-1}\right](j,i)
				 f\left(s_n  \left[\h_n\right](j,i) ,n\right)\left[\bm{\mu}_{n|n-1}\right](j,i)
				 +\sigma^2_w \I \\
& \K =   \left[\Sigma_{n|n-1}\right](j,i)
				 f\left(s_n  \left[\h_n\right](j,i) ,n\right)\left[\bm{\mu}_{n|n-1}\right](j,i)^{\top}
				 \S^{-1}	 \\			 
&\left[\bm{\mu}_{n|n}\right](j,i) = \left[\bm{\mu}_{n-1|n-1}\right](j,i) + \K \widetilde{\y}\\				 
&\left[\Sigma_{n|n}\right](j,i) =	\left(\I- \K f\left(s_n  \left[\h_n\right](j,i) ,n\right)\left[\bm{\mu}_{n|n-1}\right](j,i)\right)			 
				 \left[\Sigma_{n|n-1}\right](j,i)
\end{align*}

\label{Kalman_filter}
\end{algorithmic}
\end{algorithm}
\newpage

%%%%%%%%%%%%%%%55

\end{document}